\documentclass[12pt,twoside, a4paper]{article}
\def\pd{\partial}
\def\mc{\mathcal}
\def\ul{\underline}

\usepackage[dvips]{graphicx}
\usepackage{amssymb}
\usepackage[bbgreekl]{mathbbol}
\usepackage{amssymb,amsmath}
\usepackage{graphicx}
\usepackage{dsfont}
\usepackage{caption}
\usepackage{subcaption}
\usepackage{mathtools}
\usepackage{verbatim}
\usepackage{graphicx}
\usepackage{multirow}
\usepackage[outline]{contour}
\usepackage{xcolor,colortbl}
\input{epsf.sty}  
\begin{document}
\begin{center}
\Large{\textbf{New Janus interfaces from four-dimensional $N=3$ gauged supergravity}}
\end{center}
\vspace{1 cm}
\begin{center}
\large{\textbf{Parinya Karndumri}}
\end{center}
\begin{center}
String Theory and Supergravity Group, Department
of Physics, Faculty of Science, Chulalongkorn University, 254 Phayathai Road, Pathumwan, Bangkok 10330, Thailand
\end{center}
E-mail: parinya.ka@hotmail.com \vspace{1 cm}\\
\begin{abstract}
We construct new supersymmetric Janus solutions from four-dimensional $N=3$ gauged supergravity coupled to eight vector multiplets with $SO(3)\times SU(3)$ gauge group. This gauged supergravity admits a supersymmetric $N=3$ $AdS_4$ vacuum with the full $SO(3)\times SU(3)$ symmetry unbroken and a family of supersymmetric $N=1,2,3$ and non-supersymmetric $AdS_4$ vacua with different residual symmetries. These are expected to be dual to supersymmetric Chern-Simons-Matter (CSM) theories in three dimensions via the AdS/CFT correspondence. By considering a truncation to three complex scalar fields, we find a number of Janus solutions preserving $N=1$ supersymmetry that describes conformal interfaces within $N=3$ CSM theories with $SO(3)\times SU(3)$ and $SU(2)_{\textrm{diag}}\times U(1)$ symmetries.
\end{abstract}
\newpage
\section{Introduction}
Supersymmetric solutions of gauged supergravity in various dimensions provide a number of insights on different aspects of string/M-theory. In particular, these solutions lead to holographic descriptions of strongly coupled dynamics of superconformal field theories (SCFTs) via the AdS/CFT correspondence. Within this type of holographic solutions, $(d+1)$-dimensional Janus solutions in the form of curved domain walls with $AdS_d$ slices are one of the most studied holographic solutions. These solutions describe conformal interfaces within higher-dimensional SCFTs and can be found directly in string/M-theory or by uplifting solutions of lower-dimensional gauged supergravity to ten or eleven dimensions, see \cite{Bak_Janus}-\cite{Warner_N8_uplift} for an incomplete list. The uplifted solutions are obtained via consistent truncations of string/M-theory to the corresponding gauged supergravity of interest. However, in many cases, the embedding of gauged supergravity in higher dimensions is presently not known. Nevertheless, the study of supersymmetric solutions from gauged supergravity is still useful in the holographic context.    
\\
\indent In this paper, we are interested in constructing new supersymmetric Janus solutions from $N=3$ gauged supergravity in four dimensions. This gauged supergravity with semisimple gauge groups has been constructed long ago in \cite{N3_Ferrara}, see also \cite{N3_Ferrara2,Castellani_book}. The most interesting $N=3$ gauged supergravity is obtained by coupling the supergravity multiplet to eight vector multiplets leading to $SO(3)\times SU(3)$ gauge group. The supersymmetric $AdS_4$ vacuum preserving the full $N=3$ supersymmetry and $SO(3)\times SU(3)$ gauge symmetry is expected to describe an effective theory of eleven-dimensional supergravity compactified on a trisasakian $N^{010}$ manifold \cite{Castellani_Romans}. The complete spectrum of this truncation has been found in \cite{N3_spectrum1,N3_spectrum2}, and a possible dual SCFT in the form of a Chern-Simons-Matter (CSM) theory has been proposed in \cite{Ring_N3_superfield,Shadow_N3_multiplet}, see also \cite{N3_and_QHE} for a related holographic study of quantum Hall effect from this eleven-dimensional background. From a holographic point of view, the aforementioned $N=3$ gauged supergravity has been studied in \cite{N3_SU2_SU3}, see also \cite{N3_4D_gauging} for another work on $N=3$ gauged supergravity, in which another supersymmetric $N=3$ $AdS_4$ vacuum with $SU(2)_{\textrm{diag}}\times U(1)$ symmetry has been found. A holographic RG flow between the $SO(3)\times SU(3)$ $AdS_4$ critical point to this new critical point has also been given in \cite{N3_SU2_SU3}.    
\\
\indent In a recent work \cite{N3_4D_Mario}, $N=3$ gauged supergravity has been constructed in the duality covariant formalism using the embedding tensor. This formulation, in principle, encodes all possible supersymmetric deformations of $N=3$ supergravity coupled to vector multiplets, the only matter multiplets within $N=3$ supersymmetry. In addition, the $N=3$ gauged supergravity with $SO(3)\times SU(3)$ gauge group has been shown to admit new classes of supersymmetric and non-supersymmetric $AdS_4$ vacua with $N=0,1,2,3$ supersymmeties of which the $N=3$ $AdS_4$ vacuum with $SU(2)_{\textrm{diag}}\times U(1)$ symmetry found in \cite{N3_SU2_SU3} is a particular point on these spaces of $AdS_4$ vacua. These classes of $AdS_4$ vacua have been found by truncating the scalar manifold to $\left[SU(1,1)/U(1)\right]^3$ with three complex scalars. We are interestested in supersymmetric Janus solutions within these truncations. 
\\
\indent A number of Janus solutions from this $N=3$ gauged supergravity in $SO(2)\times SU(2)\times SO(2)$ and $SO(2)\times SO(2)\times SO(2)$ truncations has been found in \cite{N3_Janus}. However, within these truncations, there exist only Janus solutions interpolating between the $SO(3)\times SU(3)$ $AdS_4$ vacua on both sides of the interfaces. In this work, by considering different truncations studied in \cite{N3_4D_Mario}, we will see that there are Janus solutions interpolating between $N=3$ supersymmetric $AdS_4$ vacua with both $SO(3)\times SU(3)$ and $SU(2)_{\textrm{diag}}\times U(1)$ or $SO(3)_{\textrm{diag}}$ symmetries. These solutions provide first examples of Janus solutions within $N=3$ gauged supergravity that involve more than one $AdS_4$ vacuum. We expect these solutions to be useful in holographic studies of conformal interfaces in strongly coupled $N=3$ CSM theories. 
\\
\indent The paper is organized as follows. In section \ref{N3theory}, we review the matter-coupled
$N=3$ gauged supergravity in the embedding tensor formalism. We then consider electric $SO(3)\times SU(3)$ gauge group and review families of supersymmetric $AdS_4$ vacua in section \ref{SO3_SU3}. The analysis of BPS equations and the corresponding supersymmetric Janus solutions are given in section \ref{Janus_solutions}. Conclusions and comments on the results are given in section \ref{conclusions}. In the appendix, we present the explicit form of relevant BPS equations.
\section{$N=3$ gauged supergravity}\label{N3theory}
We begin with a review of $N=3$ gauged supergravity in four dimensions constructed in \cite{N3_4D_Mario} using the embedding tensor formalism. We will mainly collect relevant formulae for constructing supersymmetric Janus solutions and refer to \cite{N3_4D_Mario} and \cite{Mario_Phys_Report} for more detail.
\\
\indent The $N=3$ supergravity multiplet contains the following field content
\begin{equation}
(e^{\hat{\mu}}_\mu, \psi_{A \mu }, A^{AB}_{\mu}, \chi).
\end{equation}
These are given respectively by the graviton $e^{\hat{\mu}}_\mu$, three
gravitini $\psi_{A \mu }$, three vectors $A_{\mu A}=\frac{1}{2}\epsilon_{ABC}A^{BC}_\mu$ and one spin-$\frac{1}{2}$
field $\chi$. Indices $\mu,\nu,\ldots =0,\ldots,3$ and $\hat{\mu},\hat{\nu},\ldots=0,\ldots,3$ are respectively space-time and tangent
space indices while indices $A,B,\ldots=1,2,3$ denote the fundamental representation of $SU(3)_R\subset U(3)_R$
R-symmetry. 
\\
\indent The only matter multiplets for $N=3$ supersymmetry are given by vector multiplets, and the supergravity
multiplet can couple to an arbitrary number $n$ of vector multiplets. Each vector multiplet contains one vector field $A^\mu$, four
gaugini in a triplet $\lambda_A$ and a singlet $\lambda$ of $SU(3)_R$, and three complex scalars $z_A$. With indices $i,j,\ldots =1,\ldots, n$ labeling each of the vector multiplets, we can write the field content of the $n$ vector multiplets as  
\begin{equation}
(A^i_\mu, \lambda_{Ai}, \lambda_i, {z_A}^i)
\end{equation}
We also note that all spinors are subject to the chirality projection conditions
\begin{equation}
\psi_{\mu A}=\gamma_5\psi_{\mu A},\qquad \chi=\gamma_5\chi,\qquad
\lambda_{Ai}=\gamma_5\lambda_{Ai},\qquad
\lambda_i=-\gamma_5\lambda_i
\end{equation}
with charge-conjugate spinors having the opposite chirality.
\\
\indent The $3n$ complex or equivalently $6n$ real scalar fields are described by the coset manifold $SU(3,n)/SU(3)\times SU(n)\times U(1)$. The global symmetry of ungauged $N=3$ supergravity is given by the isometry of the scalar manifold $SU(3,n)$ of which the actions on vector fields include the electric-magnetic duality under which electric vector fields $A^\Lambda_\mu=(A^{AB}_\mu,A^i_\mu)$ together with their magnetic duals $\tilde{A}_{\Lambda \mu}$ transform. The embedding of $SU(3,n)$ in the duality group $Sp(6+2n,\mathbb{R})$ is described by the matrix ${\mc{R}[T]_M}^N$ for $T\in SU(3,n)$. The symplectic indices $M,N$ are defined as $V^M=(V^\Lambda,V_\Lambda)$ for any symplectic vector $V^M$. The matrix ${\mc{R}[T]_M}^N$ satisfy the relation
\begin{equation}
{\mc{R}[T]_M}^P{\mc{R}[T]_N}^Q\mathbb{C}_{PQ}=\mathbb{C}_{MN}
\end{equation}
for 
\begin{equation}
\mathbb{C}=\left(\begin{array}{cc}
\mathbf{0}_{3+n} & \mathbf{I}_{3+n} \\
-\mathbf{I}_{3+n} & \mathbf{0}_{3+n}
\end{array}\right).
\end{equation}
\indent The scalar fields can be parametrized by the coset representative $L$. The left-invariant one-form $\Omega$ constructed from $L$ gives the $SU(3)\times SU(n)\times U(1)$ connection $Q$ and the vielbein on $SU(3,n)/SU(3)\times SU(n)\times U(1)$ according to the relation
\begin{equation} 
\Omega=L^{-1}dL=Q+P\, .
\end{equation}
In the fundamental representation of $SU(3,n)$, we have
\begin{equation}
\Omega=\left(\begin{array}{cc}
{Q_{AB}}^{CD} & P_{ABj} \\
P^{iCD} & {Q^i}_j
\end{array}\right)
\end{equation}
with $P_{ABi}=(P^{iAB})^*=P_{iAB}$ and $P^{ABi}=(P_{ABi})^*=P^{iAB}$. We will also write
\begin{equation}
{P_A}^i=\frac{1}{2}\epsilon_{ABC}P^{BCi}=({P_i}^A)^*\, .
\end{equation}
The coset representative in the symplectic representation $\mc{R}$ will be written as
\begin{equation}
{\mc{R}[L]_M}^N={L_M}^N\, .
\end{equation} 
\indent The most general gaugings of the $N=3$ supergravity coupled to vector multiplets can be completely determined by the embedding tensor ${\Theta_M}^\alpha$ with $\alpha$ being the $SU(3,n)$ adjoint index. The gauge generators are then given by
\begin{equation}
X_M={\Theta_M}^\alpha t_\alpha
\end{equation}
with $t_\alpha$ being generators of $SU(3,n)$. In the representation $\mc{R}$, we can write
\begin{equation}
{X_{MN}}^P={\Theta_M}^\alpha {\mc{R}[t_\alpha]_N}^P
\end{equation}
which are generalized structure constants of the gauge group $G_0\subset SU(3,n)$. Consistency and supersymmetry require the embedding tensor to satisfy the following constraints
\begin{equation}
{X_{(MN}}^R\mathbb{C}_{P)R}=0\qquad \textrm{and}\qquad \left[X_M,X_N\right]=-{X_{MN}}^PX_P\, .\label{constraint}
\end{equation}
\indent In general, the group $G_0$ that is not a subgroup of the off-shell symmetry of the Lagrangian can be gauged provided that the corresponding embedding tensor satisfies the above constraints. Furthermore, the magnetic vector fields can also participate in the gauging. However, gaugings of this type require the introduction of two-form fields transforming in the adjoint representation of $SU(3,n)$. Both magnetic vector and two-form fields do not have kinetic terms but enter the gauged Lagrangian via topological terms, so they do not introduce extra degrees of freedom. 
\\
\indent In the present paper, we are only interested in electric $SO(3)\times SU(3)$ gauge group since, as pointed out in \cite{N3_4D_Mario}, the dyonic embedding of this group in $SU(3,n)$ does not lead to any new physics. Moreover, we will only consider supersymmetric Janus solutions which involve only the metric and scalar fields. Therefore, for simplicity in various expressions, we will set all the vector fields to zero from now on. With all these, the bosonic Lagrangian of $N=3$ gauged supergravity coupled to vector multiplets can be written as
\begin{equation}
e^{-1}\mc{L}=\frac{1}{2}R-{P_{\mu A}}^i{P^\mu}_i^{\phantom{i}A}-V
\end{equation}   
with $e=\sqrt{-g}$. We also note that we use mostly plus signature $(-1,1,1,1)$ for the metric as oppose to that in \cite{N3_4D_Mario}.
\\
\indent The scalar potential is given in terms of the fermion-shift matrices as
\begin{equation}
V=\frac{1}{3}\left(-12S_{AB}S^{AB}+N^AN_A+N_{Ai}N^{Ai}+{N_{iA}}^B{N^{iA}}_B\right)
\end{equation}
in which $S_{AB}$, $N_A$, $N_{Ai}$, and ${N_{iA}}^B$ can be obtained from various components of the so-called T-tensor defined by
\begin{equation}
{T_{\ul{M}\ul{N}}}^{\ul{P}}={(\mathbb{L}^{-1})_{\ul{M}}}^M {(\mathbb{L}^{-1})_{\ul{N}}}^N{\mathbb{L}_P}^{\ul{P}}{X_{MN}}^P\, .
\end{equation}
Indices $\ul{M},\ul{N},\ldots$ refer to the complex basis $V^{\ul{M}}=(V^{\ul{\Lambda}},V_{\ul{\Lambda}})$ with $V^{\ul{\Lambda}}=(V^{AB},V_i)$ and $V_{\ul{\Lambda}}=(V^{\ul{\Lambda}})^*$. This basis is more convenient for coupling to fermions. $V^{AB}=-V^{BA}$ and $V_i$ correspond respectively to $(\mathbf{3},\mathbf{1})_{-1}$ and $(\mathbf{1},\mathbf{n})_{\frac{3}{n}}$ representations of $SU(3)\times SU(n)\times U(1)$ arising from the decomposition of the fundamental representation $\mathbf{3}+\mathbf{n}$ of $SU(3,n)$ via
\begin{equation}
\mathbf{3}+\mathbf{n}\rightarrow (\mathbf{3},\mathbf{1})_{-1}\oplus (\mathbf{1},\mathbf{n})_{\frac{3}{n}}\, .
\end{equation}
This complex basis is related to the symplectic basis $V^M$ introduced earlier by the following relation
\begin{equation}
 V^M={(A^\dagger O)^M}_{\ul{N}}V^{\ul{N}}
\end{equation} 
with   
\begin{equation}   
A=\frac{1}{\sqrt{2}}\left(\begin{array}{cc}
\mathbf{I}_{3+n} & i\mathbf{I}_{3+n} \\
\mathbf{I}_{3+n} & -i\mathbf{I}_{3+n}
\end{array}\right)\qquad \textrm{and}\qquad O=\left(\begin{array}{cccc}
\mathbf{I}_{3} & \mathbf{0}_{3\times n}& \mathbf{0}_{3}& \mathbf{0}_{3\times n} \\
\mathbf{0}_{n\times 3} & \mathbf{0}_{n}& \mathbf{0}_{n\times 3}& \mathbf{I}_{n} \\
\mathbf{0}_{3} & \mathbf{0}_{3\times n}& \mathbf{I}_{3}& \mathbf{0}_{3\times n} \\
\mathbf{0}_{n\times 3} & \mathbf{I}_{n}& \mathbf{0}_{n\times 3}& \mathbf{0}_{n} 
\end{array}\right).    
\end{equation}    
We also note that, in this basis, an element ${T_{\ul{\Lambda}}}^{\ul{\Sigma}}$ of $SU(3,n)$ in the fundamental representation $\mathbf{3}+\mathbf{n}$ takes the form
\begin{equation}
{\mc{R}_c[T]_{\ul{M}}}^{\ul{N}}=\left(\begin{array}{cc}
T & \mathbf{0}_{3+n} \\
\mathbf{0}_{3+n} & T^*
\end{array}\right).
\end{equation}       
\indent The linear constraint, the first condition in \eqref{constraint}, implies that ${T_{\ul{M}\ul{N}}}^{\ul{P}}$ only have the following components 
\begin{equation}
{T_{\ul{\Lambda}\ul{\Sigma}}}^{\ul{\Gamma}}={T_{[\ul{\Lambda}\ul{\Sigma}]}}^{\ul{\Gamma}}\qquad \textrm{and}\qquad {T^{\ul{\Lambda}\ul{\Sigma}}}_{\ul{\Gamma}}={T^{[\ul{\Lambda}\ul{\Sigma}]}}_{\ul{\Gamma}}
\end{equation}
with ${T^{\ul{\Lambda}\ul{\Sigma}}}_{\ul{\Gamma}}=({T_{\ul{\Lambda}\ul{\Sigma}}}^{\ul{\Gamma}})^*$. 
\\
\indent The fermion-shift matrices are then given explicitly by
\begin{eqnarray}
S_{AB}&=&-\frac{1}{2}\epsilon_{CD(A}{T^{CD}}_{B)},\nonumber \\
N^A&=&{T^{BA}}_B,\nonumber \\
N_{Ai}&=&\epsilon_{ABC}{T^{BC}}_i,\nonumber \\
{N^{iA}}_B&=&-2{T^{iA}}_B+{\delta^A}_B{T^{iC}}_C\, .
\end{eqnarray}
Finally, supersymmetry transformations of fermions are given by
\begin{eqnarray}
\delta\psi_{A\mu}&=&D_\mu\epsilon_A-S_{AB}\gamma_\mu \epsilon^B,\nonumber \\
\delta\chi&=&N^A\epsilon_A,\nonumber \\
\delta\lambda_{Ai}&=&\epsilon_{ABC}{P_{\mu i}}^C\gamma^\mu \epsilon^B+{N_{iA}}^B\epsilon_B,\nonumber \\
\delta \lambda_i&=&{P_{\mu i}}^A\gamma^\mu \epsilon_A+N_{iA}\epsilon^A\, .\label{SUSY_trans}
\end{eqnarray}
\section{$SO(3)\times SU(3)$ gauge group and supersymmetric $AdS_4$ vacua}\label{SO3_SU3}
We now consider the case of $n=8$ vector multiplets with $SO(3)\times SU(3)$ gauge group that is electrically embedded in $SU(3,8)$ global symmetry. In this case, only the electric components of the embedding tensor ${\Theta_\Lambda}^\alpha$ are non-vanishing. The gauge generators $X^\Lambda$ coupled to magnetic vector fields vanish while the electric gauge generators in the complex basis take the form 
\begin{eqnarray}
& &\mc{R}_c[t_{SO(3)}]=\left(\begin{array}{cccc}
\textrm{Adj}(SO(3)) & \mathbf{0}_{3\times 8}& \mathbf{0}_{3}& \mathbf{0}_{3\times 8} \\
\mathbf{0}_{8\times 3} & \mathbf{0}_{8}& \mathbf{0}_{8\times 3}& \mathbf{0}_{8} \\
\mathbf{0}_{3} & \mathbf{0}_{3\times 8}& \textrm{Adj}(SO(3))& \mathbf{0}_{3\times 8} \\
\mathbf{0}_{8\times 3} & \mathbf{0}_{8}& \mathbf{0}_{8\times 3}& \mathbf{0}_{8} 
\end{array}\right),\nonumber \\
& &\mc{R}_c[t_{SU(3)}]=\left(\begin{array}{cccc}
\mathbf{0}_{3} & \mathbf{0}_{3\times 8}& \mathbf{0}_{3}& \mathbf{0}_{3\times 8} \\
\mathbf{0}_{8\times 3} & \textrm{Adj}(SU(3))& \mathbf{0}_{8\times 3}& \mathbf{0}_{8} \\
\mathbf{0}_{3} & \mathbf{0}_{3\times 8}& \mathbf{0}_{3}& \mathbf{0}_{3\times 8} \\
\mathbf{0}_{8\times 3} & \mathbf{0}_{8}& \mathbf{0}_{8\times 3}& \textrm{Adj}(SU(3)) 
\end{array}\right).
\end{eqnarray}  
Non-vanishing components of the generalized gauge structure constants are given by
\begin{equation}
{X_{\ul{\Lambda}\ul{\Sigma}}}^{\ul{\Gamma}}=(g_1\epsilon_{ABC},g_2f_{ijk})\qquad \textrm{and} \qquad X_{\ul{\Lambda}\phantom{\ul{\Gamma}}\ul{\Sigma}}^{\phantom{\ul{\Lambda}}\ul{\Gamma}}=-{X_{\ul{\Lambda}\ul{\Sigma}}}^{\ul{\Gamma}}
\end{equation}
with $f_{ijk}$ being the $SU(3)$ structure constants. The explicit form for $f_{ijk}$ can be obtained by using standard Gell-Mann matrices. $g_1$ and $g_2$ are coupling constants for the $SO(3)$ and $SU(3)$ factors, respectively.
\\
\indent Following \cite{N3_4D_Mario}, we then consider consistent truncations to three complex scalars parametrized by the scalar manifold
\begin{equation}
\left[\frac{SU(1,1)}{U(1)}\right]^3\subset SU(3,8)/SU(3)\times SU(8)\times U(1)\, . 
\end{equation}
There are two truncations considered in \cite{N3_4D_Mario} with the coset representatives in the fundamental representation of $SU(3,8)$ given by
\begin{equation}
L=e^{\mathbf{k}},\qquad \mathbf{k}=\left(\begin{array}{cc}
\mathbf{0}_{3} & \mathbf{X}_{3\times 8} \\
\mathbf{X}^\dagger_{8\times 3} & \mathbf{0}_{8}
\end{array}\right)
\end{equation}
and
\begin{eqnarray}
& &\mathbf{X}^{(i)}=\left(\begin{array}{cccccccc}
z_1 & 0  & 0 & 0 & 0 & 0 & 0 & 0\\
0 &  z_2  & 0 & 0 & 0 & 0 & 0 & 0\\
0 & 0  & z_3 & 0 & 0 & 0 & 0 & 0
\end{array}\right)\nonumber \\
& & \mathbf{X}^{(ii)}=\left(\begin{array}{cccccccc}
0 & z_1  & 0 & 0 & 0 & 0 & 0 & 0\\
0 &  0  & 0 & 0 & z_2 & 0 & 0 & 0\\
0 & 0  & 0 & 0 & 0 & 0 & z_3 & 0
\end{array}\right).
\end{eqnarray}
Each truncation gives rise to a large class of $AdS_4$ vacua with $N=0,1,2,3$ supersymmetries. 
\\
\indent Using the above coset representative with 
\begin{equation}
z_a=\phi_a e^{i\zeta_a},\qquad a=1,2,3
\end{equation}
and the coset representative in the complex basis of the form
\begin{equation}
{\mathbb{L}_{c\ul{M}}}^{\ul{N}}={\mc{R}_c[L]_{\ul{M}}}^{\ul{N}}=\left(\begin{array}{cc}
L & \mathbf{0}_{3+n} \\
\mathbf{0}_{3+n} & L^*
\end{array}\right),
\end{equation}
we can compute the T-tensor and find the scalar potential for the first truncation as follows 
\begin{eqnarray}
V^{(i)}&=&-g_1^2\left[3+2\cosh2\phi_3+\cosh2\phi_1(2+\cosh2\phi_2+\cosh2\phi_3)\right. \nonumber \\
& &\left.+\cosh2\phi_2(2+\cosh2\phi_3) \right]+g_2^2\left[3+\cosh2\phi_2(\cosh2\phi_3-2)\right. \nonumber \\
& &\left.-2\cosh2\phi_3+\cosh2\phi_1(\cosh2\phi_2+\cosh2\phi_3-2)\right]
\end{eqnarray}
while the second truncation gives 
\begin{equation}
V^{(ii)}=V^{(i)}\left(g_2\rightarrow \frac{g_2}{2}\right).
\end{equation}
Both of these potentials admit a supersymmetric $N=3$ $AdS_4$ critical point at the origin of the scalar manifold given by
\begin{equation} 
\phi_1=\phi_2=\phi_3=0,\qquad V_0=-12g_1^2,\qquad L^2=\frac{1}{4g_1^2}\, .
\end{equation}
$V_0$ is the cosmological constant, and $L$ is the $AdS_4$ radius related to $V_0$ via the relation
\begin{equation}
L^2=-\frac{3}{V_0}\, .
\end{equation} 
It should be noted that the scalar potential does not depend on all the $\zeta_a$ scalars. These scalars correspond to flat directions of the potential and are dual to marginal deformations in the dual SCFTs in three dimensions. The full $SO(3)\times SU(3)$ gauge symmetry is unbroken at this critical point.
\\
\indent In addition, $V^{(i)}$ and $V^{(ii)}$ admit another class of $AdS_4$ vacua given by
\begin{eqnarray}
\textrm{I}&:&\qquad \phi_1=\phi_2=\phi_3=\frac{1}{2}\ln\left[\frac{g_2+g_1}{g_2-g_1}\right],\qquad V_0=-\frac{12g_1^2g_2^2}{g_2^2-g_1^2},\\
\textrm{II}&:&\qquad \phi_1=\phi_2=\phi_3=\frac{1}{2}\ln\left[\frac{g_2+2g_1}{g_2-2g_1}\right],\qquad V_0=-\frac{12g_1^2g_2^2}{g_2^2-4g_1^2}\, .
\end{eqnarray}
Although the scalar potential does not depend on the phases $\zeta_a$, unbroken supersymmetry at the $AdS_4$ vacua does depend on the values of $\zeta_a$ \cite{N3_4D_Mario}. For generic values of $\zeta_a$, both critical points I and II break all supersymmetries. On the other hand, these critical points preserve some supersymmetry for particular values of $\zeta_a$ satisfying the following relations 
\begin{eqnarray}
N=1&:&\qquad \zeta_1=\zeta_2+\zeta_3,\nonumber \\
N=2&:&\qquad \zeta_1=\zeta_2\qquad \textrm{and}\qquad  \zeta_3=0,\nonumber \\
N=3&:&\qquad \zeta_1=\zeta_2=\zeta_3=0
\end{eqnarray}
modulo permutations among $\zeta_a$. Furthermore, the two classes of critical points preserve different unbroken gauge symmetries since $\mathbf{X}^{(i)}$ and $\mathbf{X}^{(ii)}$ are invariant under different subgroups of $SO(3)\times SU(3)$. In particular, critical points I and II with $N=3$ supersymmetry respectively preserve $SU(2)_{\textrm{diag}}\times U(1)$ and $SO(3)_{\textrm{diag}}$ symmetries, for more detail, see \cite{N3_4D_Mario}. 
\\
\indent Apart from different unbroken gauge symmetries, the scalar potentials and $AdS_4$ critical points from the second truncation can be obtained from the first one by substituting $g_2$ by $\frac{g_2}{2}$. Moreover, all the results of the analysis performed in the next section in truncation $ii$ can also be obtained from those in truncation $i$ by replacing $g_2$ with $\frac{g_2}{2}$. Therefore, we will only consider the first truncation in this paper to avoid an unnecessary repetition.

\section{Supersymmetric Janus solutions}\label{Janus_solutions}
In this section, we consider supersymmetric solutions within truncation $i$ described in the previous section. Holographic RG flows between the $N=3$ supersymmetric $AdS_4$ critical point at the origin to different vacua within the two classes of non-trivial $AdS_4$ critical points I and II have already been given in \cite{N3_4D_Mario}. The solutions have been found directly by solving the field equations, so these solutions also include RG flows to non-supersymmetric $AdS_4$ vacua. In this paper, we will consider supersymmetric Janus solutions in the form of $AdS_3$-sliced domain walls within truncation $i$ by solving BPS equations obtained from the vanishing of supersymmetry transformations of fermions.   

\subsection{BPS equations}
We now consider the metric ansatz of the form
\begin{equation}
ds^2=e^{2A(r)}\left(e^{\frac{2\xi}{\ell}}dx^2_{1,1}+d\xi^2\right)+dr^2
\end{equation}
which is a curved domain wall with an $AdS_3$ slice of radius $\ell$ and $dx^2_{1,1}$ being the flat Minkowski metric in two dimensions. With the following choice of vielbein  
\begin{equation}
e^{\hat{\alpha}}=e^{A+\frac{\xi}{\ell}}dx^\alpha,\quad \alpha=0,1,\qquad
e^{\hat{\xi}}=e^{A}d\xi,\qquad e^{\hat{r}}=dr,
\end{equation}
non-vanishing components of the spin connection are given by 
\begin{equation}
\omega^{\hat{\xi}}_{\phantom{\hat{\xi}}\hat{r}}=A'e^{\hat{\xi}},\qquad
\omega^{\hat{\alpha}}_{\phantom{\hat{\xi}}\hat{\xi}}=\frac{1}{\ell}e^{-A}e^{\hat{\alpha}},\qquad
\omega^{\hat{\alpha}}_{\phantom{\hat{\xi}}\hat{r}}=A'e^{\hat{\alpha}}
\end{equation}
with $'$ denoting the $r$-derivative.
\\
\indent All scalars $z_a=\phi_ae^{i\zeta_a}$ depend only on the coordinate $r$. Using the coset representative computed from $\mathbf{X}^{(i)}$, we find the kinetic term for scalars of the form
\begin{equation}
\mc{L}_{\textrm{kin}}=-\sum_{a=1}^3\left[{\phi'_a}^2+\frac{1}{4}\sinh^22\phi_a{\zeta'_a}^2\right]. 
\end{equation} 
The gravitino-shift matrix takes a diagonal form and is given by
\begin{equation}
S_{AB}=-\delta_{AB}\left[g_1\cosh\phi_1\cosh\phi_2\cosh\phi_3-g_2e^{i(\zeta_B-\zeta_C-\zeta_D)}\sinh\phi_1\sinh\phi_2\sinh\phi_3\right]\label{S_eigen}
\end{equation}
with $B\neq C\neq D$. 
\\
\indent To further analyze the BPS conditions, it is useful to introduce the ``superpotential'' defined by the eigenvalue of $S_{AB}$ along the Killing spinors
\begin{equation}
S_{AB}=-\frac{1}{2}\mc{W}_A\delta_{AB}\, .
\end{equation}
In general, the three eigenvalues given in \eqref{S_eigen} are not equal, but any of them can be used as the superpotential, as pointed out in \cite{N3_4D_Mario}. For definiteness, we will choose $\mc{W}=\mc{W}_1$ in terms of which the scalar potential can be written as
\begin{equation} 
V=2G^{rs}\frac{\pd W}{\pd \Phi^r}\frac{\pd W}{\pd \Phi^s}-3W^2
\end{equation}
with $W=|\mc{W}|$ and $\Phi^r=(\phi_1,\phi_2,\phi_3,\zeta_1,\zeta_2,\zeta_3)$, $r,s=1,2,3,\ldots 6$. Explicitly, we have 
\begin{equation}
\mc{W}=2g_1\cosh\phi_1\cosh\phi_2\cosh\phi_3-2g_2e^{i(\zeta_1-\zeta_2-\zeta_3)}\sinh\phi_1\sinh\phi_2\sinh\phi_3\, .
\end{equation}
\indent $G^{rs}$ is the inverse of the scalar metric appearing in the scalar kinetic term as
\begin{equation}
\mc{L}_{\textrm{kin}}=-\frac{1}{2}G_{rs}\pd_\mu \Phi^r\pd^\mu \Phi^s\, .
\end{equation}
Explicitly, we have $G_{rs}=(2\delta_{ab},\frac{1}{2}\sinh^22\phi_a\delta_{a+3,b+3})$, $a,b=1,2,3$. We have also verified that both truncations $i$ and $ii$ give vanishing Yang-Mills currents implying that all the vector fields can be consistently set to zero. 
\\
\indent We now consider supersymmetry transformations of fermions given in \eqref{SUSY_trans}. The analysis is essentially the same as in \cite{warner_Janus} and \cite{N3_Janus}, so we will only give the main results and refer to these references for more detail. We will use Majorana representation with all gamma matrices $\gamma^\mu$ real and $\gamma_5=i\gamma_{\hat{0}}\gamma_{\hat{1}}\gamma_{\hat{\xi}}\gamma_{\hat{r}}$
purely imaginary and $\epsilon^A=(\epsilon_A)^*$. The Killing spinor in this case is taken to be $\epsilon=\epsilon_1$. We will also write $\epsilon^*=\epsilon^1$. We begin with the gravitino variations along $x^\alpha$ which give 
\begin{equation}
A'\gamma_{\hat{r}}\epsilon+\frac{1}{\ell}e^{-A}\gamma_{\hat{\xi}}\epsilon+\mc{W}\epsilon^*=0\,
.\label{dPsi_mu_eq}
\end{equation}
Following \cite{warner_Janus}, we impose the following projectors
\begin{equation}
\gamma^{\hat{r}}\epsilon=e^{i\Lambda}\epsilon^*\qquad \textrm{and}\qquad \gamma_{\hat{\xi}}\epsilon=i\kappa e^{i\Lambda}\epsilon^*\label{Janus_proj}
\end{equation}
with a real phase $\Lambda$ and $\kappa^2=1$. The values of $\kappa=1$ or $\kappa=-1$ correspond to the chiralities of the Killing spinor on the two-dimensional interface. In the present case, these correspond to $N=(1,0)$ or $N=(0,1)$ superconformal symmetry, respectively. 
\\
\indent Taking the complex conjugate of equation \eqref{dPsi_mu_eq} and iterating lead to the following equation
\begin{equation}
A'^2=W^2-\frac{1}{\ell^2}e^{-2A}\label{dPsi_BPS_eq}
\end{equation}
which is equivalent to the integrability condition on $\epsilon$ along $x^\alpha$ and $\xi$ directions. Using the two projectors in \eqref{Janus_proj}, we find that equation \eqref{dPsi_mu_eq} gives the phase function of the form
\begin{equation}
e^{i\Lambda}=-\frac{\mc{W}}{A'+\frac{i\kappa}{\ell}e^{-A}}\,
.\label{complex_phase}
\end{equation}
\indent The equation coming from $\delta \psi_{A\hat{\xi}}=0$ takes the form
\begin{equation}
e^{-A}\pd_\xi\epsilon+\frac{1}{2}A'\gamma_{\hat{\xi}\hat{r}}\epsilon
+\frac{1}{2}\mc{W}\gamma_{\hat{\xi}}\epsilon^*=0\, .
\end{equation}
Using equation \eqref{dPsi_mu_eq}, we find
\begin{equation}
\pd_\xi\epsilon=\frac{1}{2\ell}\epsilon
\end{equation}
which leads to the solution 
\begin{equation}
\epsilon=e^{\frac{\xi}{2\ell}}\tilde{\epsilon}
\end{equation}
for $\xi$-independent $\tilde{\epsilon}$. 
\\
\indent Finally, the gravitino variation along the radial $r$ direction gives
\begin{equation}
\pd_r \epsilon+\frac{1}{2}\mc{W}e^{-i\Lambda}\epsilon=0
\end{equation}
which, by equation \eqref{dPsi_mu_eq}, can be rewritten as
\begin{equation}
\pd_r\epsilon-\frac{1}{2}A'\epsilon-\frac{i\kappa}{2\ell}e^{-A}\epsilon=0\, .\label{delta_psi_r}
\end{equation}
With all these conditions, the Killing spinor can accordingly be written as
\begin{equation}
\epsilon=e^{\frac{A}{2}+\frac{\xi}{2\ell}+i\frac{\Lambda}{2}}\varepsilon^{(0)}
\end{equation}
in which the spinor $\varepsilon^{(0)}$ satisfies
\begin{equation}
\gamma_{\hat{r}}\varepsilon^{(0)}=\varepsilon^{(0)*}\qquad
\textrm{and}\qquad
\gamma_{\hat{\xi}}\varepsilon^{(0)}=i\kappa\varepsilon^{(0)*}\,
.
\end{equation}
In general, $\varepsilon^{(0)}$ can have an $r$-dependent phase determined by \eqref{delta_psi_r}.
\\
\indent We now move to the supersymmetry transformations of spin-$\frac{1}{2}$ fields. Within the truncation under consideration here, it turns out that $N^A=0$, so $\delta \chi=0$ equation is identically satisfied. Using the $\gamma_r$ projection given in \eqref{Janus_proj}, we obtain the following equation from $\delta\lambda_i$
\begin{eqnarray}
e^{i\Lambda}(\phi_1'-i\cosh\phi_1\sinh\phi_1\zeta_1')&=&2g_1\sinh\phi_1\cosh\phi_2\cosh\phi_3\nonumber \\
& &-2g_2e^{i(\zeta_1-\zeta_2-\zeta_3)}\cosh\phi_1
\sinh\phi_2\sinh\phi_3\nonumber \\
&=&\frac{\pd \mc{W}}{\pd \phi_1}\, .
\end{eqnarray}
Similarly, from $\delta \lambda_{iA}$, we find 
\begin{eqnarray}
e^{-i\Lambda}(\phi_2'-i\cosh\phi_2\sinh\phi_2\zeta'_2)&=&\frac{\pd \mc{W}^*}{\pd \phi_2},\\
e^{-i\Lambda}(\phi_3'-i\cosh\phi_3\sinh\phi_3\zeta'_3)&=&\frac{\pd \mc{W}^*}{\pd \phi_3}\, .
\end{eqnarray}
Using the phase function from \eqref{complex_phase}, we can solve all these equations and find the BPS equations for scalar fields. These equations are given by
\begin{eqnarray}
\phi_1'&=&-2\frac{A'}{W}\frac{\pd W}{\pd\phi_1}-4\textrm{csch}2\phi_1\frac{\kappa e^{-A}}{\ell W}\frac{\pd W}{\pd \zeta_1},\nonumber \\
\zeta_1'&=&-8\textrm{csch}^22\phi_1\frac{A'}{W}\frac{\pd W}{\pd\zeta_1}+4\textrm{csch}2\phi_1\frac{\kappa e^{-A}}{\ell W}\frac{\pd W}{\pd \phi_1},\nonumber \\
\phi_2'&=&-2\frac{A'}{W}\frac{\pd W}{\pd\phi_2}+4\textrm{csch}2\phi_2\frac{\kappa e^{-A}}{\ell W}\frac{\pd W}{\pd \zeta_2},\nonumber \\
\zeta_2'&=&-8\textrm{csch}^22\phi_2\frac{A'}{W}\frac{\pd W}{\pd\zeta_2}-4\textrm{csch}2\phi_2\frac{\kappa e^{-A}}{\ell W}\frac{\pd W}{\pd \phi_2},\nonumber \\
\phi_3'&=&-2\frac{A'}{W}\frac{\pd W}{\pd\phi_3}+4\textrm{csch}2\phi_3\frac{\kappa e^{-A}}{\ell W}\frac{\pd W}{\pd \zeta_3},\nonumber \\
\zeta_3'&=&-8\textrm{csch}^22\phi_3\frac{A'}{W}\frac{\pd W}{\pd\zeta_3}-4\textrm{csch}2\phi_3\frac{\kappa e^{-A}}{\ell W}\frac{\pd W}{\pd \phi_3}
\end{eqnarray}
with 
\begin{eqnarray}
W^2&=&4g_1^2\cosh^2\phi_1\cosh^2\phi_2\cosh^2\phi_3+4g_2^2\sinh^2\phi_1\sinh^2\phi_2\sinh^2\phi_3\nonumber \\
& &-g_1g_2\cos(\zeta_1-\zeta_2-\zeta_3)\sinh2\phi_1\sinh2\phi_2\sinh2\phi_3\, .
\end{eqnarray}
We give the explicit form of these equations in the appendix. It can also be verified that these BPS equations together with \eqref{dPsi_BPS_eq} are compatible with the second-order field equations. A similar result can be found in truncation $ii$ by setting $g_2\rightarrow \frac{g_2}{2}$. 

\subsection{Numerical Janus solutions}
\begin{figure}
         \centering
          \begin{subfigure}[b]{0.45\textwidth}
                 \includegraphics[width=\textwidth]{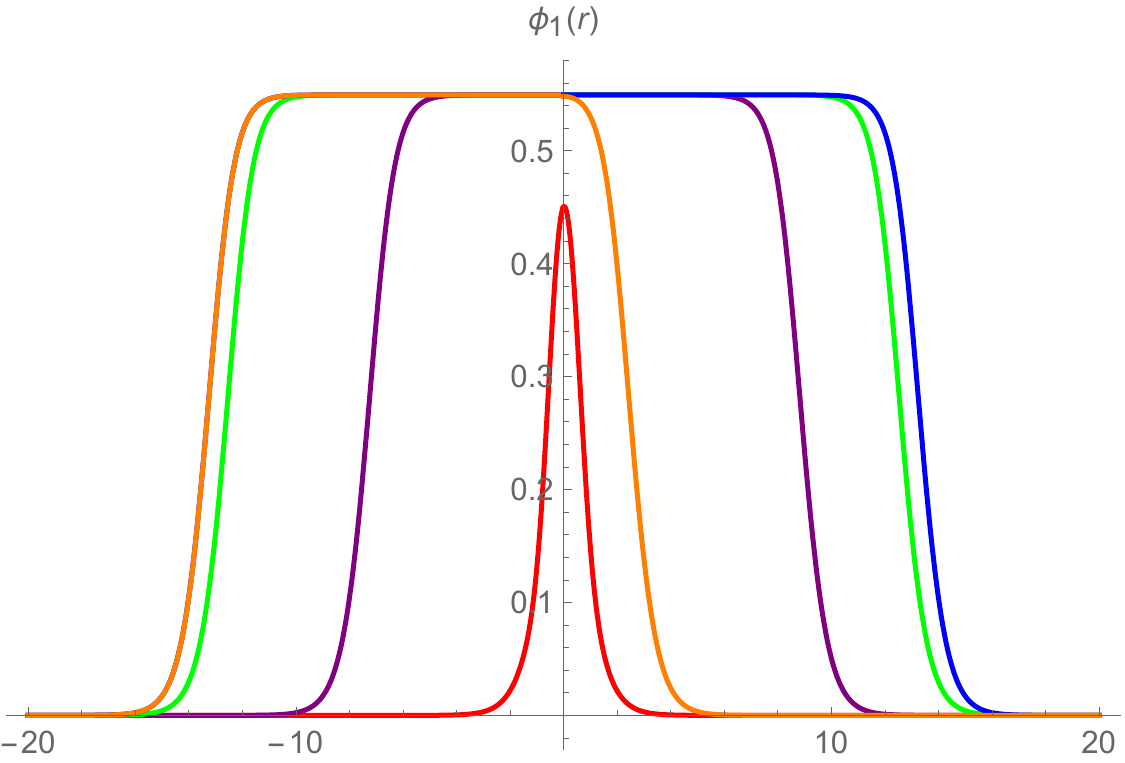}
                 \caption{Solutions for $\phi_1(r)$}
         \end{subfigure}
          \begin{subfigure}[b]{0.45\textwidth}
                 \includegraphics[width=\textwidth]{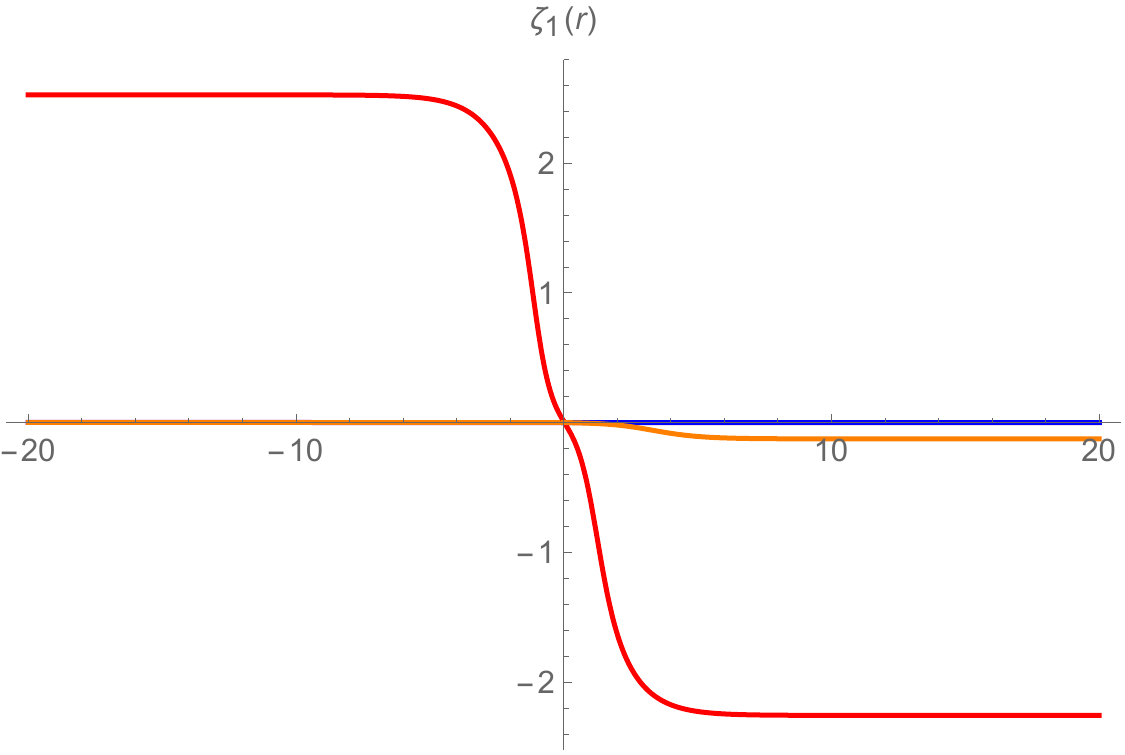}
                 \caption{Solutions for $\zeta_1(r)$}
         \end{subfigure}\\
         \begin{subfigure}[b]{0.45\textwidth}
                 \includegraphics[width=\textwidth]{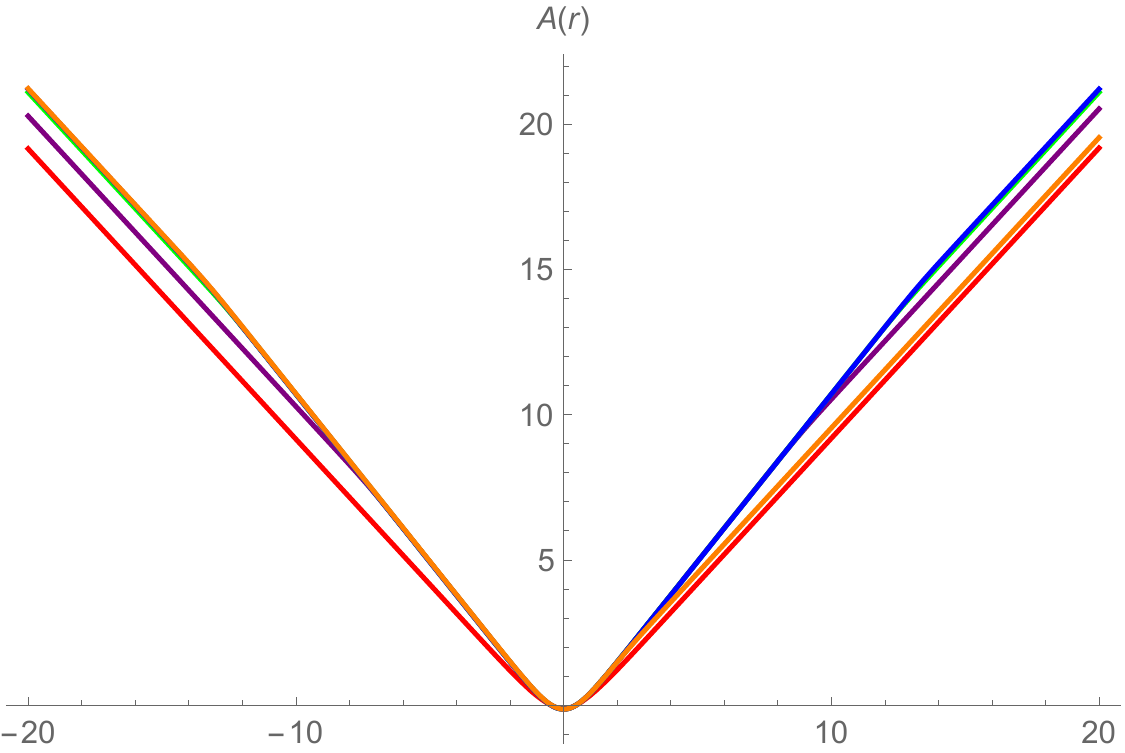}
                 \caption{Solutions for $A(r)$}
         \end{subfigure}
          \begin{subfigure}[b]{0.45\textwidth}
                 \includegraphics[width=\textwidth]{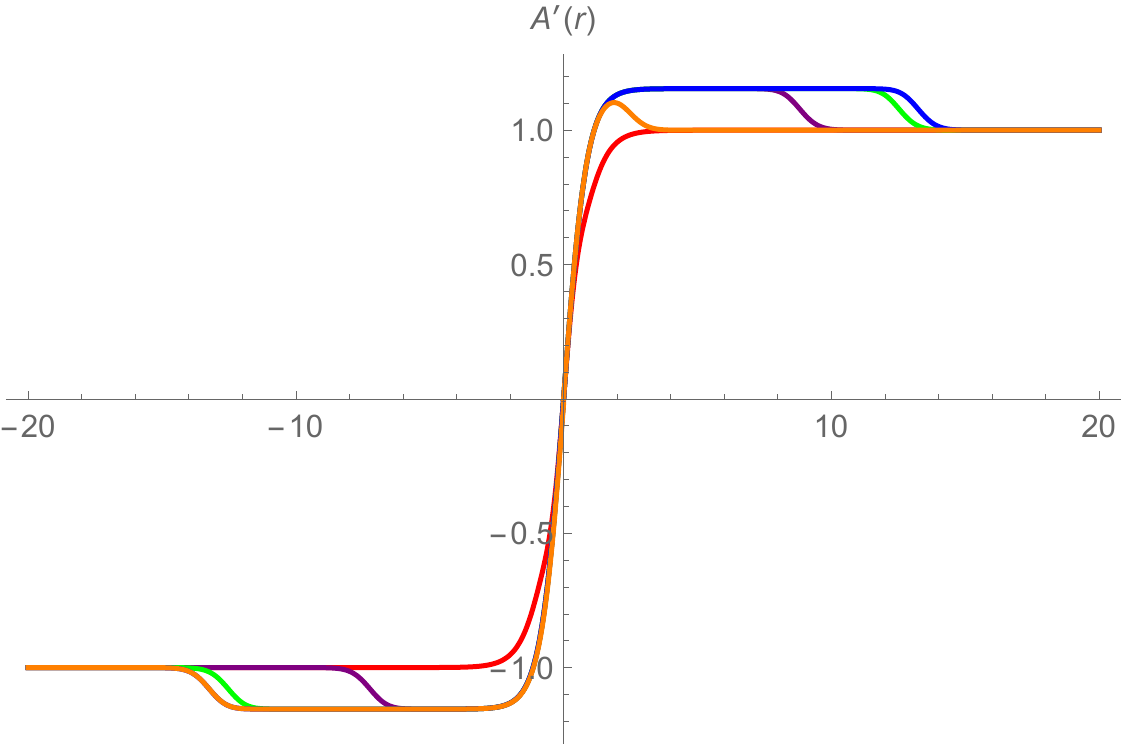}
                 \caption{Solutions for $A'(r)$}
         \end{subfigure}
\caption{Supersymmetric $N=1$ Janus solutions with $SO(2)_{\textrm{diag}}\times U(1)$ symmetry interpolating among $N=3$ supersymmetric $AdS_4$ vacua with $SO(3)\times SU(3)$ and $SU(2)_{\textrm{diag}}\times U(1)$ symmetries within a subtruncation to one complex scalar.}\label{fig1}
 \end{figure}

The resulting BPS equations are rather complicated to solve for any analytic solutions. Accordingly, we will look for numerical solutions. Unlike the RG flow solutions considered in \cite{N3_4D_Mario} in which the phase scalars $\zeta_a$ vanish, there are six active scalars in the present case. We will look at different subtruncations and consider the full set of BPS equations at the end of this section.
\\
\indent We first consider a truncation to one independent complex scalar by setting 
\begin{equation}
\phi_1=\phi_2=\phi_3\qquad \textrm{and}\qquad \zeta_2=\zeta_3=-\zeta_1\, .
\end{equation}
The BPS equations then reduce to 
\begin{eqnarray}
\phi_1'&=&-\frac{\sinh2\phi_1}{\ell W^2}\left[\ell A'(4g_1^2\cosh^4\phi_1+4g_2^2\sinh^4\phi_1-g_1g_2\cos3\zeta_1\sinh4\phi_1) \right.\nonumber \\
& & \left. +2e^{-A}g_1g_2\sin3\zeta_1\sinh2\phi_1\right],\\
\zeta_1'&=&-\frac{e^{-A}}{\ell W^2}\left[4(g_2^2-g_1^2)\cosh2\phi_1-(g_1^2+g_2^2)(3+\cosh4\phi_1) \right. \nonumber \\
& &\left.+2g_1g_2\cos3\zeta_1\sinh4\phi_1+4g_1g_2e^AA'\ell \sin3\zeta_1\sinh2\phi_1 \right]
\end{eqnarray}
together with
\begin{equation}
 {A'}^2+\frac{1}{\ell^2}e^{-2A}=W^2
\end{equation}
for
\begin{equation}
W^2=4g_1^2\cosh^6\phi_1+4g_2^2\sinh^6\phi_1-g_1g_2\cos3\zeta_1\sinh^32\phi_1\, .
\end{equation}
\indent In the above equations, we immediately see that for $\zeta_1=0$, we need to set $\ell\rightarrow \infty$. Therefore, there are no Janus solutions in the absence of pseudoscalars corresponding to imaginary parts of the complex scalars. This is in accordance with other previous works on four-dimensional Janus solutions given in \cite{warner_Janus,N3_Janus,tri-sasakian-flow,orbifold_flow}. It is also useful to point out that setting $\zeta_1=\zeta_2=\zeta_3$ does not lead to a consistent set of BPS equations unless $\zeta_1=\zeta_2=\zeta_3=0$. In particular, the BPS equations imply $\zeta'_1=-\zeta'_2$ for $\zeta_1=\zeta_2=\zeta_3$. This is also in agreement with the result of \cite{N3_Janus} in which the non-existence of Janus solutions with $SO(3)_{\textrm{diag}}\times U(1)$ symmetry, corresponding to $z_1=z_2=z_3$, has been pointed out.
\\
\indent We are now in a position to give numerical Janus solutions to the above equations. In the numerical analysis, we will choose the following numerical values of various parameters 
\begin{equation}
\kappa=-1,\qquad \ell=1,\qquad g_1=\frac{1}{2},\qquad g_2=1\, .\label{parameter}
\end{equation}
To find the solutions, we use the boundary conditions given by the values of $A=A^{(0)}$, $\phi_1=\phi_1^{(0)}$ and $\zeta_1=\zeta_1^{(0)}$ at the turning point of $A$ at which $A'(r_0)=0$ for a particular value of the radial coordinate $r=r_0$. We can also choose $r_0=0$. As in other cases, most of the initial values of $(A^{(0)},\phi_1^{(0)},\zeta_1^{(0)})$ generally lead to singular solutions approaching a singular geometry at either $r>0$ or $r<0$ side or both sides of the turning point. In this paper, we are only interested in regular solutions interpolating between $AdS_4$ vacua on both side of the interface. We find examples of these Janus solutions as shown in figure \ref{fig1}.  

\begin{figure}
         \centering
               \begin{subfigure}[b]{0.45\textwidth}
                 \includegraphics[width=\textwidth]{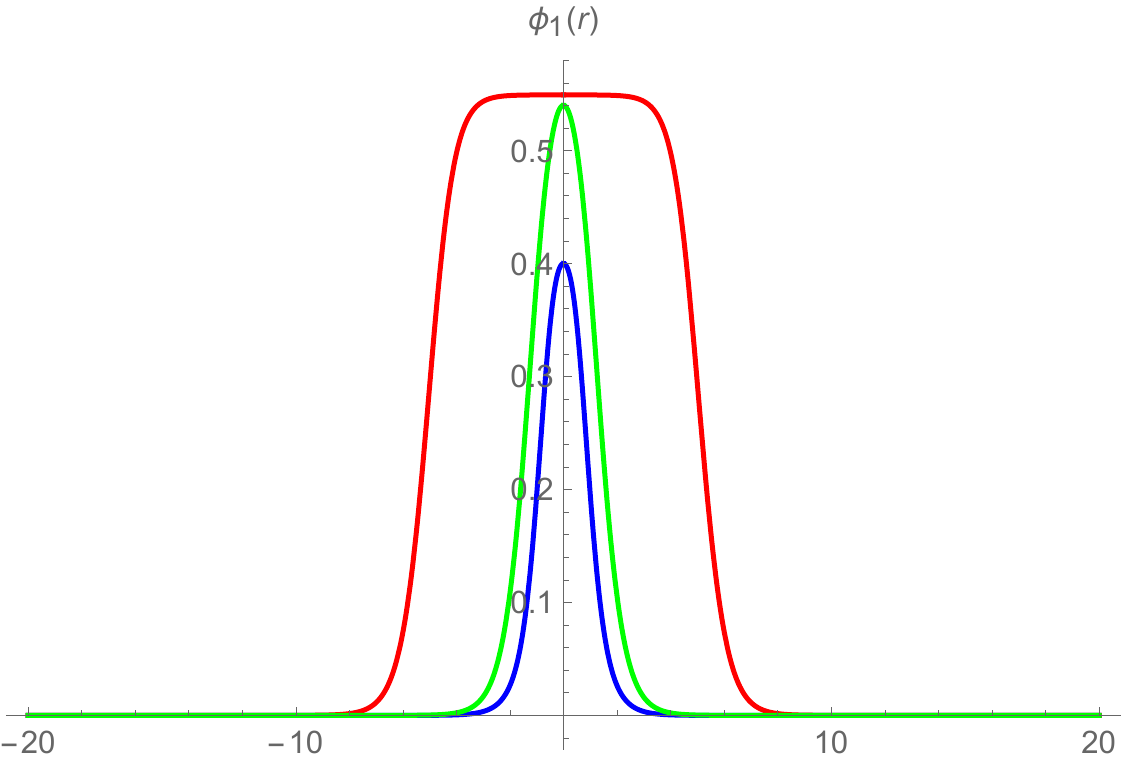}
                 \caption{Solutions for $\phi_1(r)$}
         \end{subfigure}
         \begin{subfigure}[b]{0.45\textwidth}
                 \includegraphics[width=\textwidth]{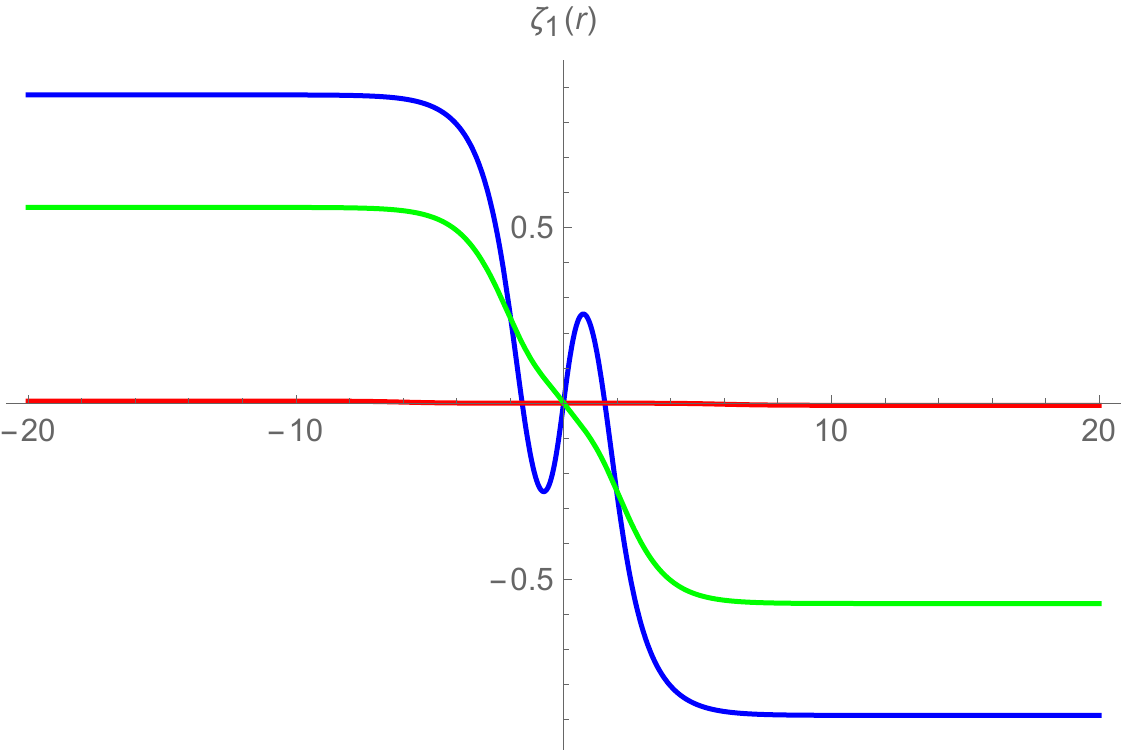}
                 \caption{Solutions for $\zeta_1(r)$}
         \end{subfigure}\\
          \begin{subfigure}[b]{0.45\textwidth}
                 \includegraphics[width=\textwidth]{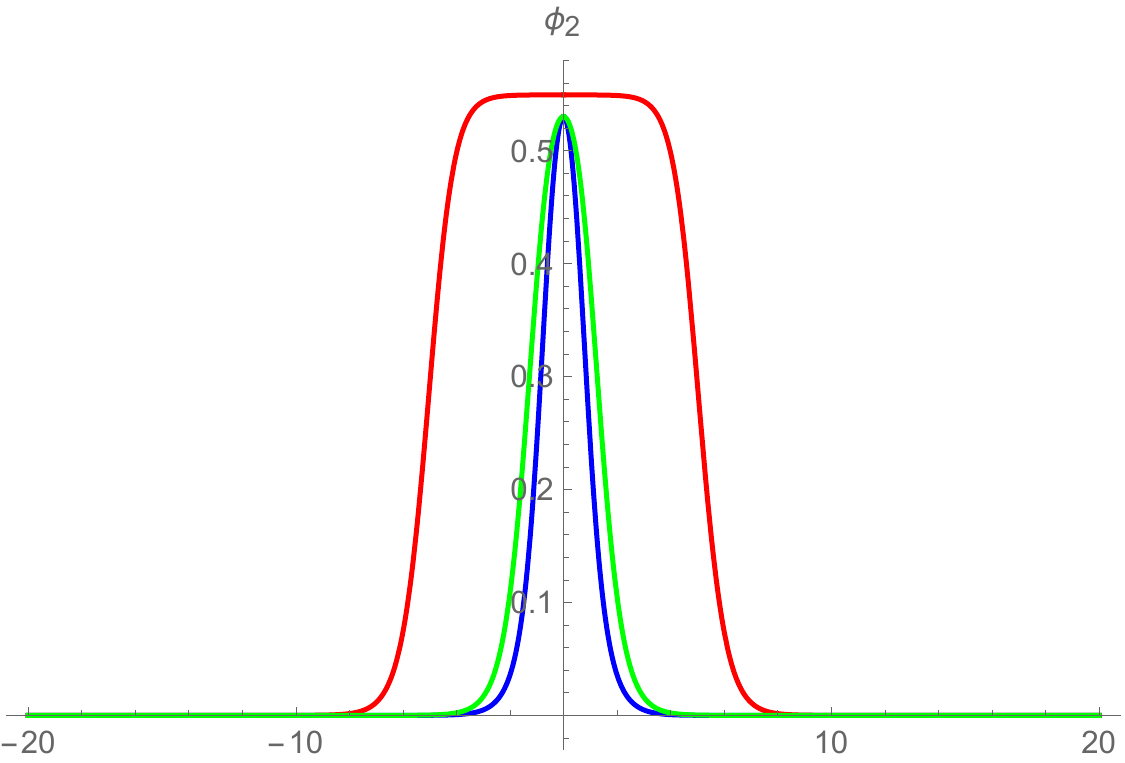}
                 \caption{Solutions for $\phi_2(r)$}
         \end{subfigure}
          \begin{subfigure}[b]{0.45\textwidth}
                 \includegraphics[width=\textwidth]{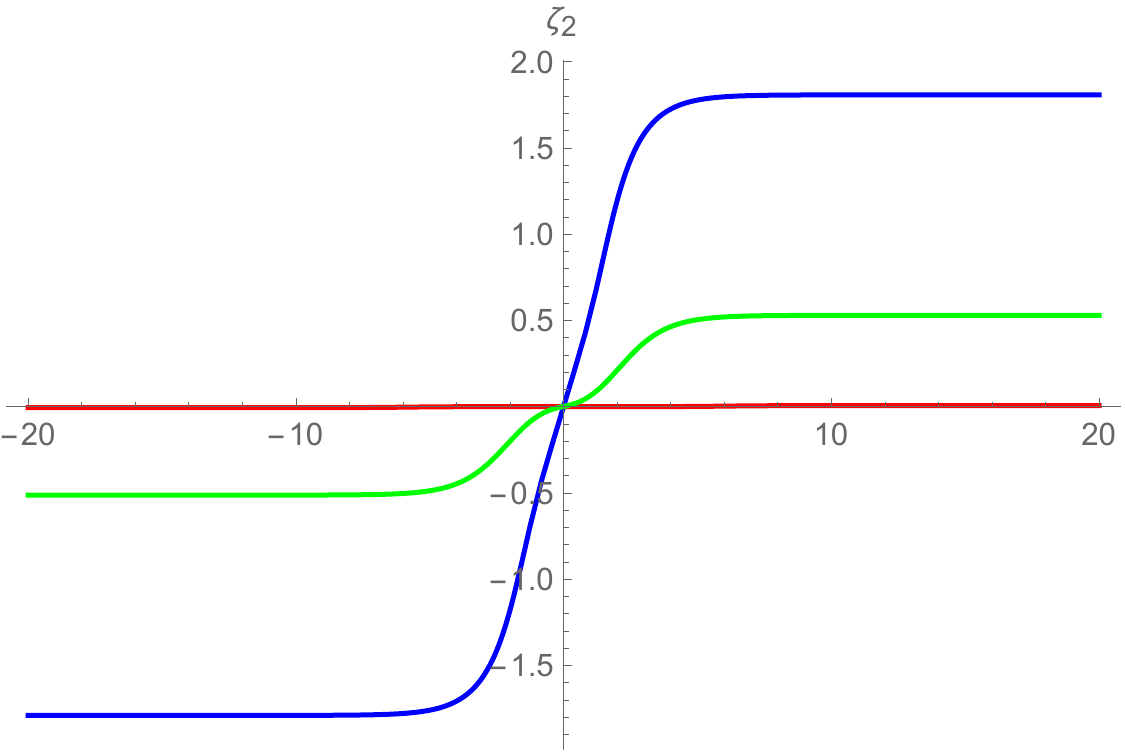}
                 \caption{Solutions for $\zeta_2(r)$}
         \end{subfigure}\\
         \begin{subfigure}[b]{0.45\textwidth}
                 \includegraphics[width=\textwidth]{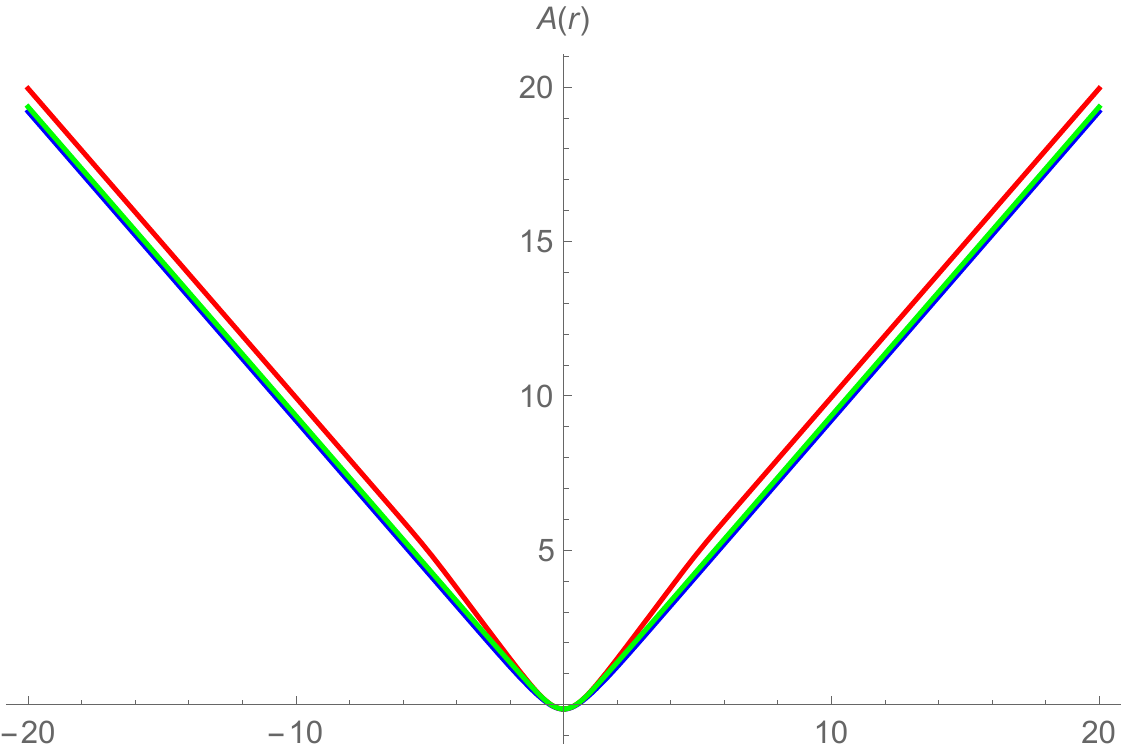}
                 \caption{Solutions for $A(r)$}
         \end{subfigure}
          \begin{subfigure}[b]{0.45\textwidth}
                 \includegraphics[width=\textwidth]{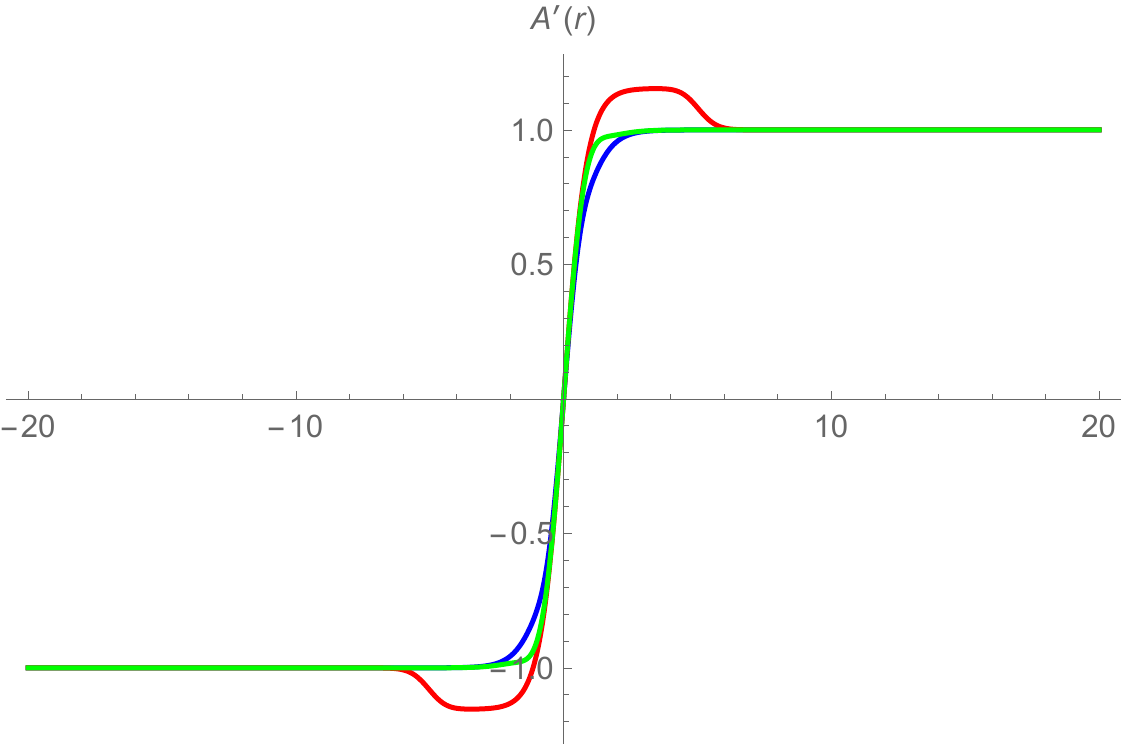}
                 \caption{Solutions for $A'(r)$}
         \end{subfigure}
\caption{Supersymmetric $N=1$ Janus solutions with $SO(2)_{\textrm{diag}}\times U(1)$ symmetry interpolating among $N=3$ supersymmetric $AdS_4$ vacua with $SO(3)\times SU(3)$ and $SU(2)_{\textrm{diag}}\times U(1)$ symmetries within a subtruncation to two complex scalars.}\label{fig2}
 \end{figure}

There are solutions that interpolate between $SO(3)\times SU(3)$ symmetric $AdS_4$ vacuum as shown by the red line in the figure. By fine-tuning the values of $(A^{(0)},\phi_1^{(0)},\zeta_1^{(0)})$, we can find solutions that approach $AdS_4$ vacuum I with $SU(2)_{\textrm{diag}}\times U(1)$ symmetry on both sides; purple, green and blue lines. For non-vanishing $\zeta_1$, the solution is invariant under $SO(2)_{\textrm{diag}}\times U(1)\subset SU(2)_{\textrm{diag}}\times U(1)$ symmetry and preserves only $N=1$ supersymmetry along the flow. During the transition between $AdS_4$ critical points on each side of the interface, $\zeta_1$ vanishes implying that the $AdS_4$ critical point on both side preserves $N=3$ supersymmetry and $SU(2)_{\textrm{diag}}\times U(1)$ symmetry. On each side of the interface, the inital $SO(3)\times SU(3)$ phase undergoes a holographic RG flow to the  $SU(2)_{\textrm{diag}}\times U(1)$ phase. These solutions give holographic descriptions of two-dimensional conformal interfaces within $N=3$ CSM theories in three dimensions. 
\\
\indent By further tuning the values of $(A^{(0)},\phi_1^{(0)},\zeta_1^{(0)})$, we also find solutions that approach $SO(3)\times SU(3)$ critical point on one side and $SU(2)_{\textrm{diag}}\times U(1)$ critical point on the other side. An example of these solutions is given by the orange line in the figure. This would provide an example of RG-flow interfaces within $N=3$ CSM theories. We end the discussion of this subtruncation by considering an asymptotic expansion of scalar fields near the $AdS_4$ critical points. At the $SO(3)\times SU(3)$ critical point, we find that the linearized BPS equations give
\begin{eqnarray}
& & \phi_1\sim \phi_0 z-\frac{\phi_0L^2_{\textrm{I}}}{\ell^2}z^3+\ldots ,\\
& &\zeta_1\sim -\frac{2L_{\textrm{I}}}{\ell}z+\frac{2\phi_0g_2L^2_{\textrm{I}}}{\ell}z^2+\ldots
\end{eqnarray}
with $\phi_0$ being a constant and $\ldots$ representing higher order terms. The coordinate $z$ is defined by $z=e^{-2g_1r}=e^{-\frac{r}{L_{\textrm{I}}}}$ for $L_{\textrm{I}}=\frac{1}{2g_1}$. To give some holographic interpretations, we consider real and imaginary parts of the complex scalar $z_1$ 
\begin{eqnarray}
& &\textrm{Re}\, z_1=\phi_1\cos\zeta_1\sim z+z^3+\ldots ,\nonumber \\
& &\textrm{Im}\, z_1=\phi_1\sin\zeta_1\sim z^2+z^3+\ldots ,
\end{eqnarray}
The standard holographic interpretation identifies the scalar Re$z_1$ and the pseudoscalar Im$z_1$ with the dual operators in the form of bosonic and fermionic bilinears of dimensions $\Delta=1$ and $\Delta=2$, respectively. From the above expansion, we find that both the boson and fermion mass terms dual to Re$z_1$ and Im$z_1$ are not turned on. We then expect the interfaces to arise from position-dependent expectation values.  
\\
\indent Similarly, the expansion near the $SU(2)_{\textrm{diag}}\times U(1)$ critical point gives 
\begin{eqnarray}
& &\textrm{Re}\, z_1\sim z^{-1}+\ldots,\nonumber \\ 
& &\textrm{Im}\, z_1\sim z^2+\ldots 
\end{eqnarray} 
with in this case $z=e^{-\frac{r}{L_{\textrm{II}}}}$ for $L_{\textrm{II}}=\frac{\sqrt{g_2^2-g_1^2}}{2g_1g_2}$. From the mass spectrum given in \cite{N3_SU2_SU3} and \cite{N3_4D_Mario}, we see that the scalar Re$z_1$ is dual to an irrelevant operator of dimension $\Delta=4$. The leading term in the expansion of Re$z_1$ indicates that this $AdS_4$ vacuum is a repulsive critical point, so finding solutions that approach this critical point needs fine-tuning in such a way that the source of this operator corresponding to $z^{-1}$ term vanishes. 
\\
\indent We then consider another subtruncation with two independent complex scalar fields obtained by setting 
\begin{equation}
\zeta_3=\zeta_2\qquad \textrm{and}\qquad \phi_3=\phi_2\, .
\end{equation}
With the same numerical values of parameters given in \eqref{parameter}, we find a number of solutions shown in figure \ref{fig2}. In this case, it is more difficult to find the suitable boundary conditions due to a bigger set of initial values for the four scalars. Similar to the previous subtruncation, there are solutions interpolating between $SO(3)\times SU(3)$ $AdS_4$ critical points and between $N=3$ $SU(2)_{\textrm{diag}}\times U(1)$ critical points arising from holographic RG flows from the $SO(3)\times SU(3)$ phases on both sides. The solutions in this case also preserve $N=1$ supersymmetry and $SO(2)_{\textrm{diag}}\times U(1)$ symmetry as in the previous case. Apart from two additional scalars, the overall structure of the resulting solutions is qualitatively the same.      

\begin{figure}
         \centering
         \begin{subfigure}[b]{0.45\textwidth}
                 \includegraphics[width=\textwidth]{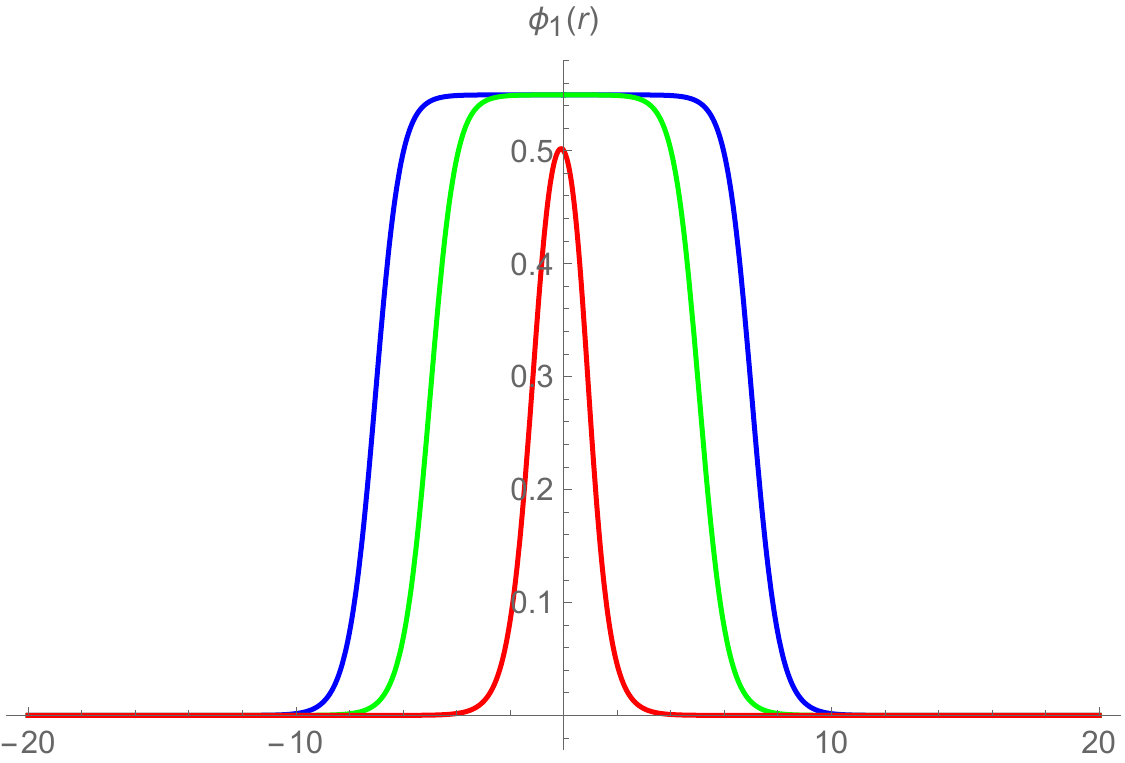}
                 \caption{Solutions for $\phi_1(r)$}
         \end{subfigure}
         \begin{subfigure}[b]{0.45\textwidth}
                 \includegraphics[width=\textwidth]{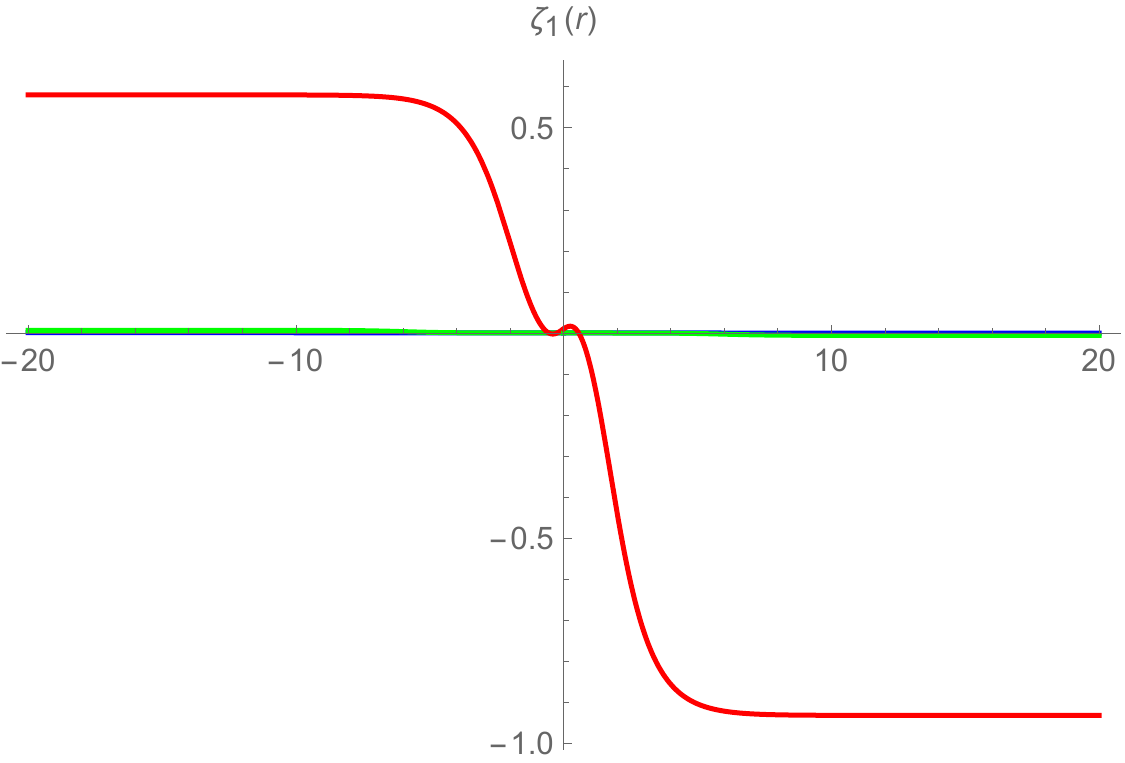}
                 \caption{Solutions for $\zeta_1(r)$}
         \end{subfigure}\\
               \begin{subfigure}[b]{0.45\textwidth}
                 \includegraphics[width=\textwidth]{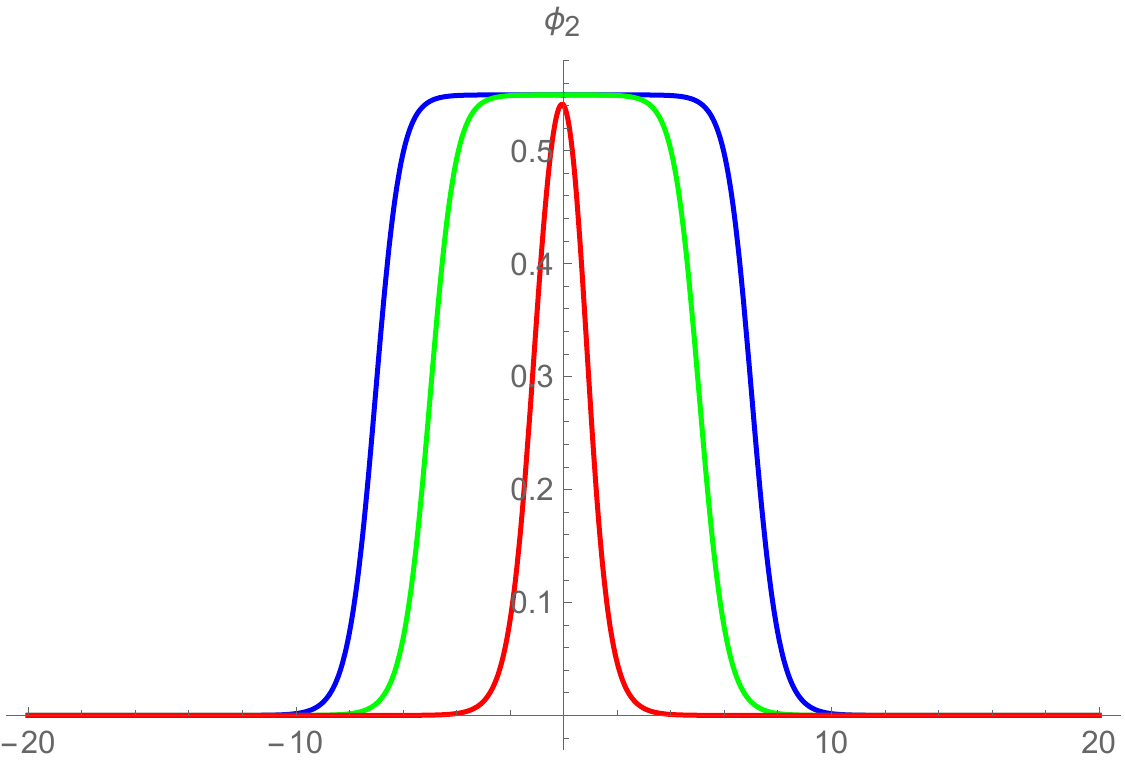}
                 \caption{Solutions for $\phi_2(r)$}
         \end{subfigure}
         \begin{subfigure}[b]{0.45\textwidth}
                 \includegraphics[width=\textwidth]{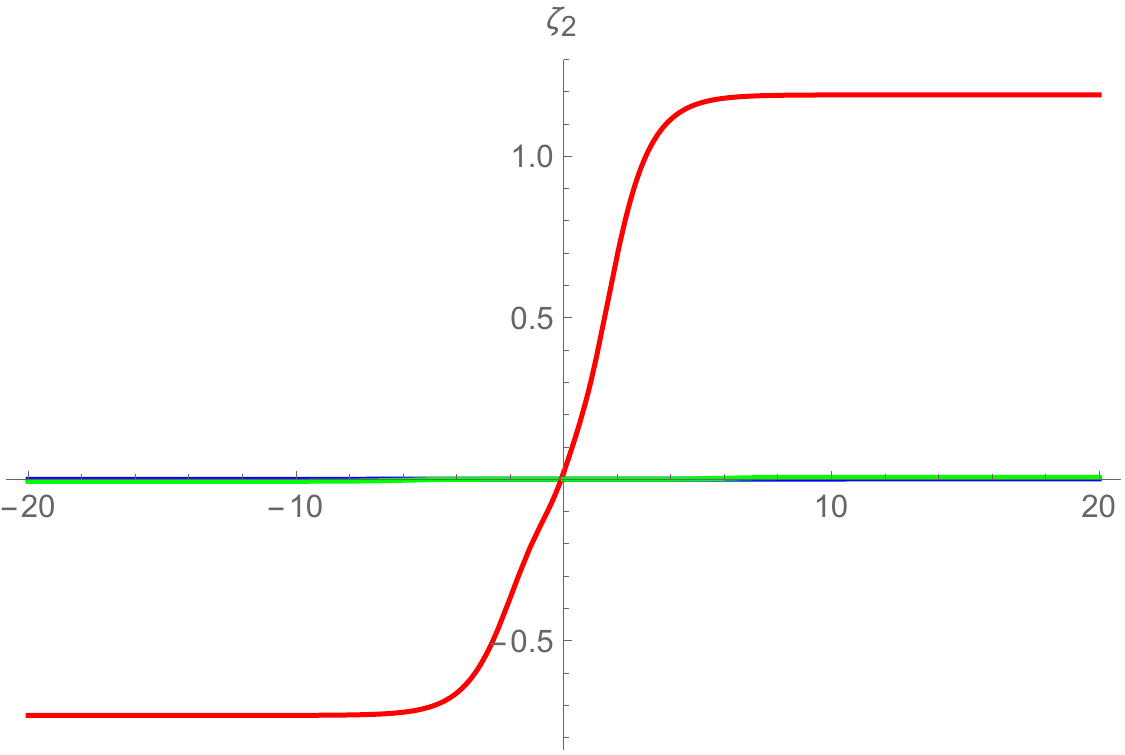}
                 \caption{Solutions for $\zeta_2(r)$}
         \end{subfigure}\\
          \begin{subfigure}[b]{0.45\textwidth}
                 \includegraphics[width=\textwidth]{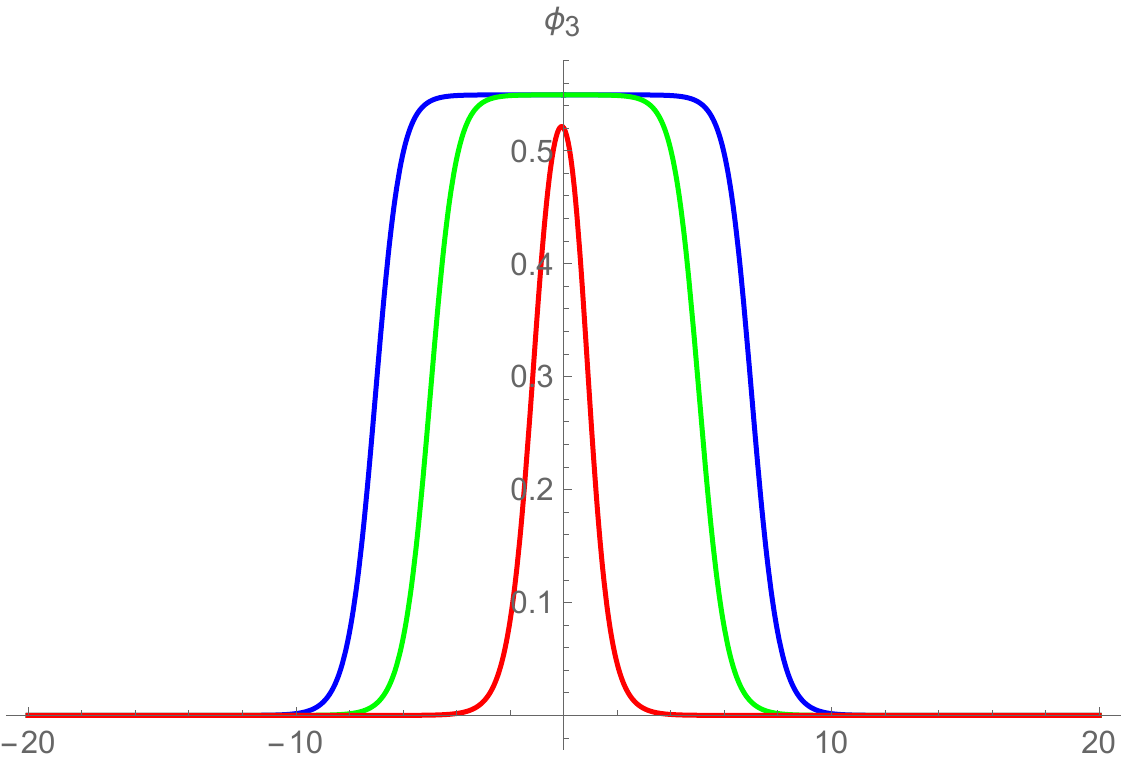}
                 \caption{Solutions for $\phi_3(r)$}
         \end{subfigure}
          \begin{subfigure}[b]{0.45\textwidth}
                 \includegraphics[width=\textwidth]{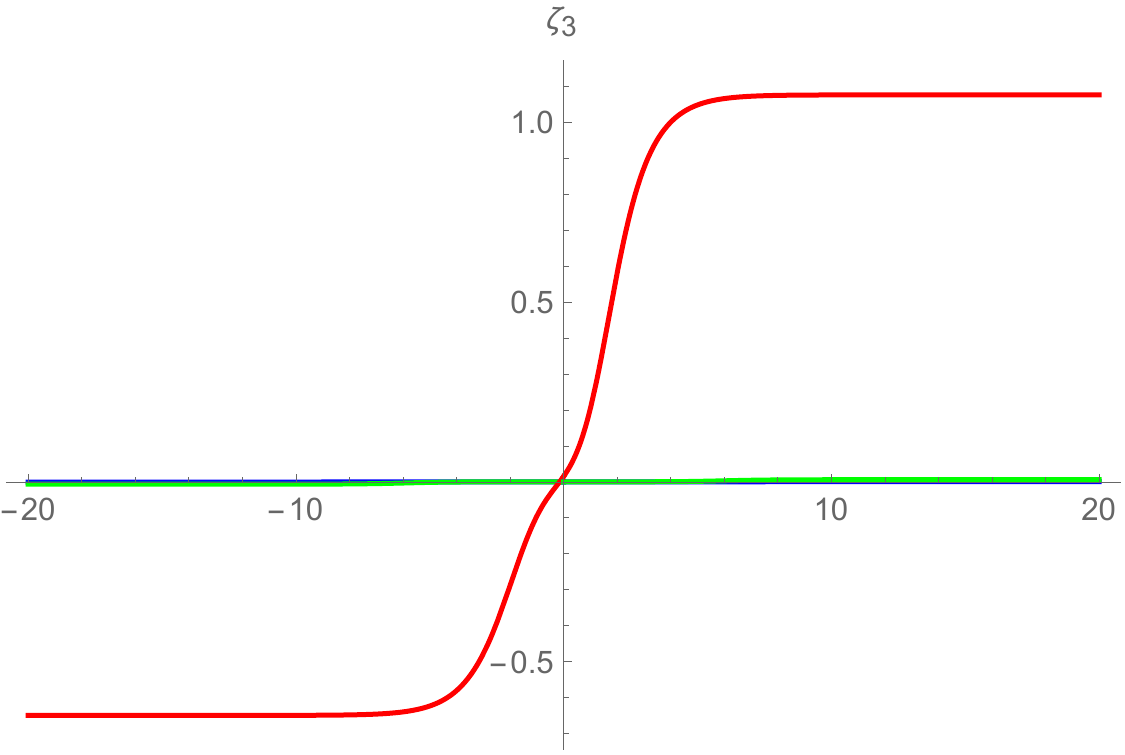}
                 \caption{Solutions for $\zeta_3(r)$}
         \end{subfigure}\\
         \begin{subfigure}[b]{0.45\textwidth}
                 \includegraphics[width=\textwidth]{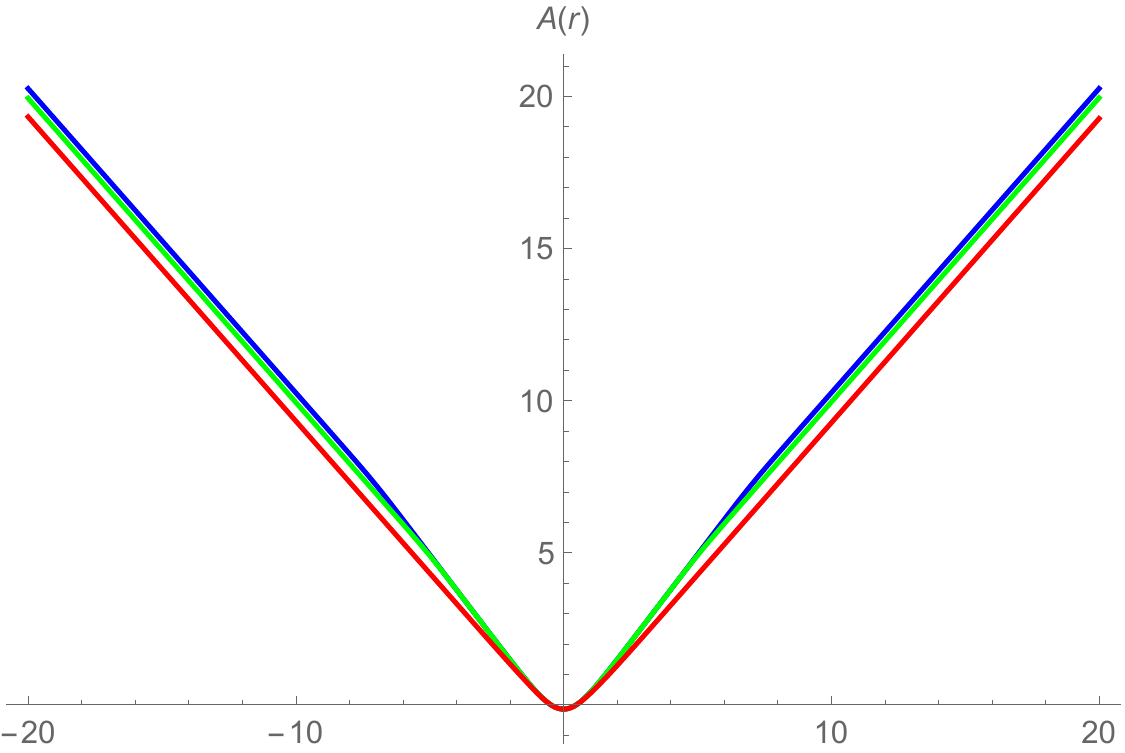}
                 \caption{Solutions for $A(r)$}
         \end{subfigure}
          \begin{subfigure}[b]{0.45\textwidth}
                 \includegraphics[width=\textwidth]{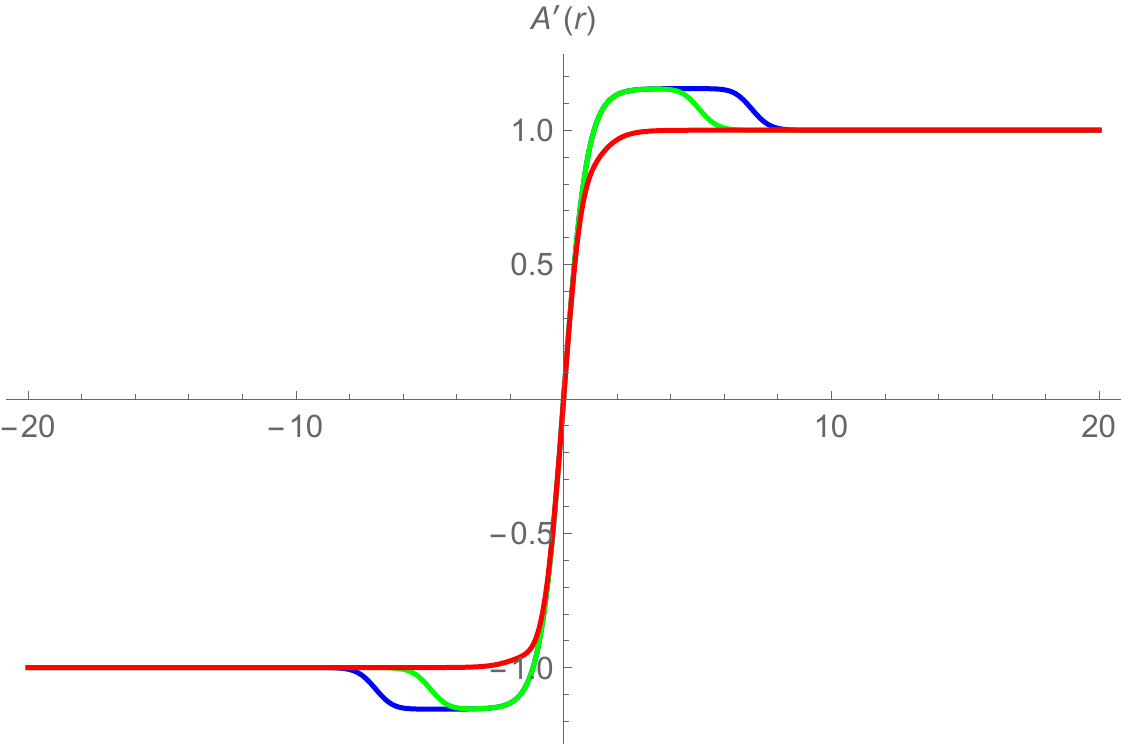}
                 \caption{Solutions for $A'(r)$}
         \end{subfigure}
\caption{Supersymmetric $N=1$ Janus solutions with $U(1)$ symmetry interpolating among $N=3$ supersymmetric $AdS_4$ vacua with $SO(3)\times SU(3)$ and $SU(2)_{\textrm{diag}}\times U(1)$ symmetries.}\label{fig3}
 \end{figure}

\indent Finally, we consider the full set of BPS equations with all three complex scalars non-vanishing. By an intensive search for suitable choices of the initial values of scalars and the warp factor $A(r)$ at the turning point, we find examples of regular Janus solutions as shown in figure \ref{fig3}. In this case, the solutions again preserve $N=1$ supersymmetry but only $U(1)$ symmetry. As can be seen from all the three figures, the general structure of solutions from the numerical analysis is similar for solutions with one, two or three complex scalars. In particular, we have not found solutions with $\zeta_a$ non-vanishing at the turning points. Accordingly, we have not found solutions describing interfaces beween $N=1$ or $N=2$ supersymmetric $AdS_4$ critical points. It is not clear whether these solutions exist or not.  
 
As previously mentioned, supersymmetric Janus solutions from truncations $ii$ can be similarly obtained by changing $g_2$ to $\frac{g_2}{2}$ in the BPS equations. In this case, the non-trivial $N=3$ $AdS_4$ critical points on each side of the interfaces preserve only a smaller $SO(3)_{\textrm{diag}}$ symmetry. 
\section{Conclusions}\label{conclusions}
In this paper, we have constructed new supersymmetric Janus solutions in $N=3$ gauged supergravity coupled to eight vector multiplets with $SO(3)\times SU(3)$ gauge group. By considering a truncation to three complex scalars, we have found a number of Janus solutions interpolating between $N=3$ $AdS_4$ vacua with $SO(3)\times SU(3)$ and $SU(2)_{\textrm{diag}}\times U(1)$ symmetries. We have given $N=1$ supersymmetric solutions preserving $SO(2)_{\textrm{diag}}\times U(1)$ symmetry with one and two independent complex scalars non-vanishing. The solutions with all three complex scalars non-vanishing is $N=1$ supersymmetric and invariant only under $U(1)\subset SO(3)\times SU(3)$ and a discrete quaternionic group \cite{N3_4D_Mario}. These solutions should be holographically dual to various two-dimensional conformal interfaces within $N=3$ CSM theories in three dimensions. 
\\
\indent It would be interesting to identify the conformal interfaces dual to the supergravity solutions found in this paper in the framework of three-dimensional superconformal field theories. It could also be of particular interest to find possible embedding of the $N=3$ $SO(3)\times SU(3)$ gauged supergravity in ten or eleven dimensions. This would allow us to uplift Janus solutions found in this paper and \cite{N3_Janus} and holographic RG flows studied in \cite{N3_SU2_SU3,N3_4D_Mario} to string/M-theory in which the complete holography can be worked out. This might be achieved by an extension or generalization of the truncation given in \cite{N010_truncation_Cassani}. Using the duality covariant formulation of $N=3$ gauged supergravity constructed in \cite{N3_4D_Mario} to study other gauge groups and possible symplectic deformations along the line of \cite{Inverso_symplectic} is also interesting. Finally, finding other solutions such as $AdS_4$ black holes from this $N=3$ gauged supergravity is also worth considering. We leave these issues to future works.     
\vspace{0.5cm}\\
{\large{\textbf{Acknowledgement}}} \\
This work is funded by National Research Council of Thailand (NRCT) and Chulalongkorn University under grant N42A650263. 
\appendix
\section{BPS equations}
In this appendix, we give the explicit form of the BPS equations for obtaining Janus solutions within truncation $i$. These equations are given by
\begin{eqnarray}
\phi'_1&=&-\frac{1}{\ell W^2}\left[2g_1g_2e^{-A}\sin(\zeta_1-\zeta_2-\zeta_3)\sinh2\phi_2\sinh2\phi_3 \right.\nonumber \\
& &+2\ell A'\left[2\sinh2\phi_1(g_1^2\cosh^2\phi_2\cosh^2\phi_3+g_2^2\sinh^2\phi_2\sinh^2\phi_3) \right.\nonumber \\
& &\left. \left. -g_1g_2\cos(\zeta_1-\zeta_2-\zeta_3)\cosh2\phi_1\sinh2\phi_2\sinh2\phi_3\right]\right],\\
\zeta'_1&=&-\frac{4e^{-A}}{\ell \sinh2\phi_1W^2}\left[g_1g_2\cos(\zeta_1-\zeta_2-\zeta_3)\cosh2\phi_1\sinh2\phi_2\sinh2\phi_3\right.\nonumber \\
& & -2\sinh2\phi_1(g_1^2\cosh^2\phi_2\cosh^2\phi_3+g_2^2\sinh^2\phi_2\sinh^2\phi_3)\nonumber \\
& &\left. g_1g_2A'\ell e^A\sin(\zeta_1-\zeta_2-\zeta_3)\sinh2\phi_2\sinh2\phi_3\right],\\
\phi'_2&=&-\frac{1}{\ell W^2}\left[2g_1g_2e^{-A}\sin(\zeta_1-\zeta_2-\zeta_3)\sinh2\phi_1\sinh2\phi_3 \right.\nonumber \\
& &+\ell A'\left[2\sinh2\phi_2(g_1^2\cosh^2\phi_1\cosh^2\phi_3+g_2^2\sinh^2\phi_1\sinh^2\phi_3) \right.\nonumber \\
& &\left.\left. -g_1g_2\cos(\zeta_1-\zeta_2-\zeta_3)\cosh2\phi_2\sinh2\phi_1\sinh2\phi_3\right]\right],\\
\zeta'_2&=&\frac{4e^{-A}}{\ell \sinh2\phi_2W^2}\left[g_1g_2\cos(\zeta_1-\zeta_2-\zeta_3)\cosh2\phi_2\sinh2\phi_1\sinh2\phi_3\right.\nonumber \\
& & -2\sinh2\phi_2(g_1^2\cosh^2\phi_1\cosh^2\phi_3+g_2^2\sinh^2\phi_1\sinh^2\phi_3)\nonumber \\
& &\left. g_1g_2A'\ell e^A\sin(\zeta_1-\zeta_2-\zeta_3)\sinh2\phi_1\sinh2\phi_3\right],
\end{eqnarray}
\begin{eqnarray}
\phi'_3&=&-\frac{1}{\ell W^2}\left[2g_1g_2e^{-A}\sin(\zeta_1-\zeta_2-\zeta_3)\sinh2\phi_1\sinh2\phi_2 \right.\nonumber\\
& & 2\ell A'\left[2\sinh2\phi_3(g_1^2\cosh^2\phi_1\cosh^2\phi_2+g_2^2\sinh^2\phi_1\sinh^2\phi_2) \right.\nonumber \\
& &\left. \left.-g_1g_2\cos(\zeta_1-\zeta_2-\zeta_3)\cosh2\phi_3\sinh2\phi_1\sinh2\phi_2 \right]\right],
\\
\zeta'_3&=&\frac{4e^{-A}}{\ell \sinh2\phi_3W^2}\left[g_1g_2\cos(\zeta_1-\zeta_2-\zeta_3)\cosh2\phi_3\sinh2\phi_1\sinh2\phi_2\right.\nonumber \\
& & -2\sinh2\phi_3(g_1^2\cosh^2\phi_1\cosh^2\phi_2+g_2^2\sinh^2\phi_1\sinh^2\phi_2)\nonumber \\
& &\left. g_1g_2A'\ell e^A\sin(\zeta_1-\zeta_2-\zeta_3)\sinh2\phi_1\sinh2\phi_2\right].
\end{eqnarray}
The BPS equation for the warp fator $A(r)$ is given in equation \eqref{dPsi_BPS_eq}. Similar equations for truncation $ii$ can be found by replacing $g_2$ with $\frac{g_2}{2}$.

\end{document}